\documentclass[%
 reprint,
superscriptaddress,
noshowpacs,
nofootinbib,
 amsmath,amssymb,
 aps
 prc,
]{revtex4-1}

\usepackage[%
  colorlinks=true,
  urlcolor=blue,
  linkcolor=blue,
  citecolor=blue
]{hyperref}

\usepackage[normalem]{ulem} % only for the draft (strike through fonts with \sout

\usepackage{multirow}

\usepackage{graphicx}% Include figure files
\usepackage{dcolumn}% Align table columns on decimal point
\usepackage{xcolor}
\usepackage{bm}% bold math

\usepackage{tikz}
\usepackage{pgfplots}

\usepackage{mathrsfs} %mathscr fonts

\begin{document}

\preprint{APS/123-QED}

\title{Energy dependence of the prompt $\gamma$-ray emission from the (d,p)-induced fission of $^{234}$U* and $^{240}$Pu*}% Force line breaks with \\

\author{S.J. Rose} 
\thanks{Corresponding author: sunniva.rose@fys.uio.no}
\affiliation{Department of Physics, University of Oslo, 0316 Oslo, Norway}
\author{F. Zeiser}
\thanks{Corresponding author: fabio.zeiser@fys.uio.no}
\affiliation{Department of Physics, University of Oslo, 0316 Oslo, Norway}

\author{J.N. Wilson}
\affiliation{Institut de Physique Nucl\'eaire d'Orsay, CNRS/ Univ. Paris-Sud, Universit\'e Paris Saclay, 91406 Orsay Cedex, France}

\author{A. Oberstedt}
\affiliation{Extreme Light Infrastructure - Nuclear Physics (ELI-NP) / Horia Hulubei National Institute for Physics and Nuclear Engineering (IFIN-HH), 077125 Bucharest-Magurele, Romania}
\author{S. Oberstedt}
\affiliation{European Commission, Joint Research Centre, Directorate for Nuclear Safety and Security, Unit G.2 – Standards for Nuclear Safety, Security and Safeguards, 2440 Geel, Belgium}

\author{S. Siem}
\affiliation{Department of Physics, University of Oslo, 0316 Oslo, Norway}
\author{G.M. Tveten}
\affiliation{Department of Physics, University of Oslo, 0316 Oslo, Norway}

\author{L.A. Bernstein}
\affiliation{Lawrence Berkeley National Laboratory, Berkeley, CA 94720, USA}
\affiliation{University of California - Berkeley Dept. of Nuclear Engineering, Berkeley CA 94720, USA}

\author{D.L. Bleuel}
\affiliation{Lawrence Livermore National Laboratory, Livermore, CA 94551, USA}

\author{J.A. Brown}
\affiliation{Lawrence Berkeley National Laboratory, Berkeley, CA 94720, USA}

\author{L. Crespo Campo}
\affiliation{Department of Physics, University of Oslo, 0316 Oslo, Norway}

\author{F. Giacoppo}
\thanks{Current affiliation: a) Helmholtz Institute Mainz, 55099 Mainz, Germany
b) GSI Helmholtzzentrum f{\"u}r Schwerionenforschung, 64291 Darmstadt, Germany}
\affiliation{Department of Physics, University of Oslo, 0316 Oslo, Norway}%Lines break automatically or can be forced with \\

\author{A. G{\"o}rgen}
\author{M. Guttormsen}
\author{K. Hady\'nska}
\author{A. Hafreager}
\author{T.W. Hagen}
\author{M. Klintefjord}
\affiliation{Department of Physics, University of Oslo, 0316 Oslo, Norway}%Lines break automatically or can be forced with \\

\author{T.A. Laplace}
\affiliation{University of California - Berkeley Dept. of Nuclear Engineering, Berkeley CA 94720, USA}%Lines break automatically or can be forced with \\
\affiliation{Lawrence Livermore National Laboratory, Livermore, CA 94551, USA}

\author{A.C. Larsen}
\affiliation{Department of Physics, University of Oslo, 0316 Oslo, Norway}%Lines break automatically or can be forced with \\
\author{T. Renstr{\o}m}
\author{E. Sahin}
\affiliation{Department of Physics, University of Oslo, 0316 Oslo, Norway}%Lines break automatically or can be forced with \\

\author{C. Schmitt}
\affiliation{Grand Acc\'el\'erateur National d'Ions Lourd, Bd Henri Becquerel, BP 55027 - 14076 CAEN Cedex 05, France}

\author{T.G. Tornyi}
\affiliation{Department of Physics, University of Oslo, 0316 Oslo, Norway}%Lines break automatically or can be forced with \\

\author{M. Wiedeking}
\affiliation{iThemba LABS, P.O. Box 722, 7129 Somerset West, South Africa}

\date{\today}% It is always \today, today,

\begin{abstract}
Prompt fission $\gamma$-rays are responsible for approximately 5\% of the total energy released in fission, and therefore important to understand when modelling nuclear reactors.
In this work we present prompt $\gamma$-ray emission characteristics in fission, for the first time as a function of the nuclear excitation energy of the fissioning system. Emitted $\gamma$-ray spectra were measured, and $\gamma$-ray multiplicities and average and total $\gamma$ energies per fission were determined for the $^{233}$U(d,pf) reaction for excitation energies between 4.8 and 10 MeV, and for the $^{239}$Pu(d,pf) reaction between 4.5 and 9 MeV. The spectral characteristics show no significant change as a function of excitation energy above the fission barrier, despite the fact that an extra $\sim$5 MeV of energy is potentially available in the excited fragments for $\gamma$-decay. The measured results are compared to model calculations made for prompt $\gamma$-ray emission with the fission model code GEF. Further comparison with previously obtained results from thermal neutron induced fission is made to characterize possible differences arising from using the surrogate (d,p) reaction. 
\begin{description}
\item[PACS numbers]
25.85.Ge 24.75.+i 07.85.Nc
 
\end{description}
\end{abstract}

\pacs{Valid PACS appear here}% PACS, the Physics and Astronomy
\maketitle

%\tableofcontents
\section{Introduction}
Nuclear fission was discovered some 70 years ago \citep{Meitner1939,Hahn1939,Bohr1939}, but still there remain some intriguing mysteries about this complex process.
One of the least measured parts of the energy that is released in fission is the contribution that is carried away via prompt $\gamma$-ray emission. This accounts for roughly 8 MeV \citep{pleasonton1972prompt, bowman1958prompt}, or around 5\% 
of the total energy released in fission. In addition, prompt energy is dissipated via the Coulomb repulsion of the fragments, and the emission of prompt neutrons. Prompt fission $\gamma$-rays (PFG) are emitted, typically within a few nanoseconds of scission of the fragments; about 70\% of the prompt PFGs are emitted within 60 ps \citep{albinsson1971decay}, and about 95\% within 3 ns \citep{Talou2016}. PFGs are one of the least understood parts of the fission process \citep{stetcu2014properties}.

The investigation of PFG emission addresses questions in nuclear structure and reaction physics. One question deals with the de-excitation of nuclei through the emission of neutrons and $\gamma$-rays. The theoretical description of the de-excitation of neutron-rich isotopes, as being produced in neutron-induced fission, shows significant deficits in describing the neutron and $\gamma$-ray spectral shape \citep{stetcu2014properties}. To some extent this deficiency seems to be related to a limited understanding of the competing process of prompt neutron and $\gamma$ emission. Prompt fission $\gamma$-ray spectral (PFGS) data, measured as a function of excitation energy of the compound nucleus may provide important information to benchmark different models, allowing eventually arrival at a consistent description of prompt fission neutron and $\gamma$-ray emission. Furthermore, PFGs are certainly among the most sensitive observables for studying angular momentum generation in fission \citep{stetcu2014properties, schmitt1984angular}.

Understanding the PFG emission is not only useful for complete modelling of the fission process, but it also has some important practical applications for nuclear reactors.  In recent years, requests for more accurate PFGS data have motivated a series of measurements to obtain new precise values of the $\gamma$-ray multiplicities and mean photon energy release per fission in the thermal-neutron induced fission of $^{235}$U \cite{Oberstedt2013, Chyzh2013} and $^{239}$Pu \cite{Gatera2017, Chyzh2013}. With the development of advanced Generation-IV nuclear reactors, the need of new PFGS data becomes important. Since four out of six contemplated Generation-IV reactors require a fast-neutron spectrum, a wider range of incident neutron energies has to be considered \citep{Pioro2016}. Modelling of the geometrical distribution of $\gamma$-heating, in and around the reactor core, shows local deviations up to 28\% as compared to measured heat distributions, whereas accuracy within 7.5\% is mandatory \cite{NEAHPPu3}. These deviations remain mainly, despite experiment campaigns in the 1970s \citep{Pelle1971, pleasonton1972prompt, verbinski1973prompt, pleasonton1973prompt}, due to the uncertainties on the existing PFGS data \cite{Salvatores2008, billnert2013new, Oberstedt2013}. For $^{240}$Pu*, this work also responds to the high-priority request published through the OECD/NEA \citep{NEAHPPu3}.
 
In this paper we report about the first measurements of PFG emission from $^{234}$U* in the $^{233}$U(d,pf) reaction, and $^{240}$Pu* in the $^{239}$Pu(d,pf) reaction. Both target nuclei represent the fissile key nuclei for the thorium/uranium and uranium/plutonium fuel cycles, respectively. The (d,pf) reaction serves hereby as a surrogate for the neutron induced fission \citep{cramer1970neutron}. Charged-particle induced reactions allow measurements of fission observables for isotopes not easily accessible to neutron beam experiments, or for excitation energies below the neutron binding energy. They also facilitate the study of PFG characteristics as a function of compound nucleus excitation energy. We study the dependence of PFG characteristics on compound nucleus excitation energy, and possible differences between surrogate and neutron-induced fission reactions.

\section{Experimental details}
Two experiments, denoted by (A) and (B), were carried out at the Oslo Cyclotron Laboratory (OCL) of the University of Oslo, using deuteron beams, delivered by a MC-35 Scanditronix cyclotron. The $\gamma$-detector array CACTUS \citep{guttormsen1990statistical} together with the SiRi charged particle detectors \citep{guttormsen2011siri} and the NIFF detector \citep{tornyi2014new} were used to detect triple coincident events of a proton, one of the two fission fragments (FF) and $\gamma$-rays.

Experiment (A) utilized a 12.5 MeV beam incident on a $^{233}$U target, and experiment (B) had a 12 MeV beam on a $^{239}$Pu target (detailed target specifications are listed in Table \ref{tab:table1}). The targets were cleaned from decay products and other chemical impurities with an anion-exchange resin column procedure \citep{Henderson2011}, and then electroplated on a backing made of $^{9}$Be.

\begin{table}
\caption{\label{tab:table1}%
Target and beam characteristics as used in this work. Fission barrier heights are taken from Ref. \citep{capote2009ripl} }
\begin{ruledtabular}
\begin{tabular}{ccc}
Target & $^{233}$U (A)& $^{239}$Pu (B)\\
\hline
Chemical composition& metallic & metallic \\
Active diameter & 1 cm & 1 cm \\
$^{9}$Be backing (mg/cm$^{2}$) & 2.3  &  1.8 \\
Total area density (mg/cm$^{2}$) & 0.2 & 0.4 \\
Reaction & (d,pf) & (d,pf) \\
Beam energy (MeV) & 12.5 & 12 \\
Inner fission barrier, $B_\mathrm{F,a}$ (MeV) & 4.80 & 6.05\\
Outer fission barrier, $B_\mathrm{F,b}$ (MeV) & 5.50 & 5.15\\

\end{tabular}
\end{ruledtabular}
\end{table}

\begin{figure}
\centering
\includegraphics[width=0.5\textwidth]{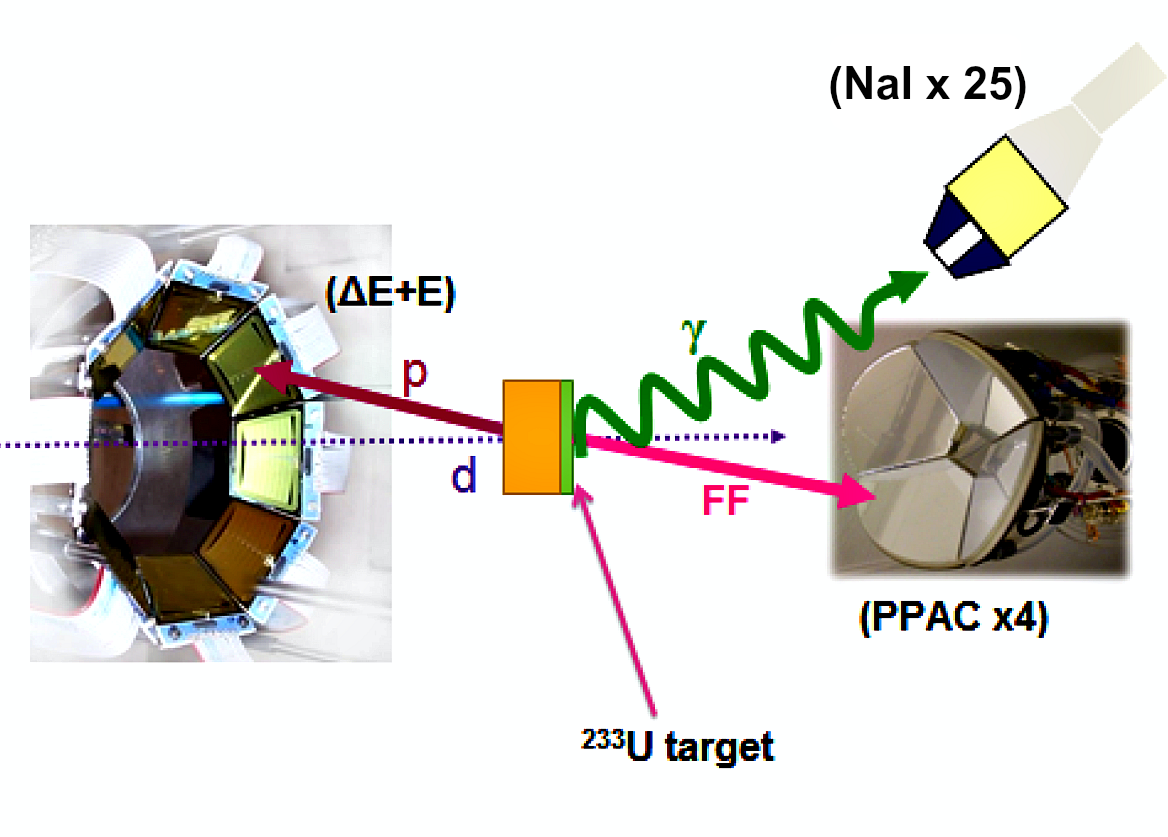}
\caption{(Color online) Schematic view of the experimental set-up for experiment (A), showing the SiRi ($\Delta$E+E) telescope, and the NIFF (PPAC) detectors, inside the reaction chamber, surrounded by the CACTUS NaI array. SiRi measures the energy of the outgoing charged-particles; NIFF detects fission fragments (FF), and CACTUS detects $\gamma$-rays – all in coincidence, within a time interval of 20 ns. The $^{233}$U target (0.2 mg/cm$^{2}$, green), on the $^{9}$Be backing (2.3 mg/cm$^{2}$, orange) was facing NIFF, and SiRi was in the backward direction relative to the beam direction (dotted, purple arrow). The setup for the $^{239}$Pu experiment was identical, except for CACTUS having 26 crystals instead of 25.}
\label{fig:experiment}
\end{figure}

For these particular experiments, the SiRi detectors were mounted in backward direction, and the NIFF detectors in forward direction, relative to the beam direction (see Fig. \ref{fig:experiment}). This setup was chosen for several reasons: Due to the thick beryllium backing, the targets had to face NIFF, to enable detection of any fission events, thereby also avoiding FF in the SiRi detector. However, the light, outgoing particles could easily penetrate the beryllium, and be detected in SiRi. Backward direction of SiRi also reduces the intensity of the elastic peak, and minimizes the exposure to protons from deuteron breakup in the target. SiRi was covered by a 21 $\mu$m thick aluminum foil, to attenuate $\delta$-electrons in the telescopes.
 
SiRi consists of 64 $\Delta$E (front) and 8 E (back) silicon detectors with thicknesses of 130 $\mu$m and 1550 $\mu$m, respectively. The detectors cover eight angles from $\theta \simeq$ 126$^{\circ}$ to 140$^{\circ}$ relative to the beam axis, in a lampshade geometry facing the target at a distance of 5 cm at an angle of 133$^{\circ}$. The total solid angle coverage is about 9\%  of 4$\pi$. In experiment (A) twenty-five, and in experiment (B) twenty-six, 12.7 cm $\times$ 12.7 cm ($5^{\prime\prime}\times 5^{\prime\prime}$) NaI(Tl) crystals were mounted on the spherical CACTUS frame, 22 cm away from the target. At a $\gamma$-ray energy of 1.33~MeV, the crystals detect $\gamma$-rays with a total efficiency of 13.6(1)\% (A), and 14.2(1)\% (B), respectively. In order to reduce the amount of Compton scattering, the detectors were collimated with lead cones. 
NIFF, consisting of four Parallel Plate Avalanche Counters (PPAC), covering 41\% of 2$\pi$, were used for tagging of fission events. For this, it is sufficient to detect one of the two fission fragments, which are emitted back-to-back.
The PPACs are placed at an angle of 45$^{\circ}$ with respect to the beam 
axis, at a distance of about 5 cm from the centre of the target.
Taking into account angular anisotropy effects in the center-of-mass 
system, Ref. \citep{ducasse2015study} found a total efficiency of about 48\%.
The particle and fission detectors were mounted in the reaction chamber, surrounded by the CACTUS array (Fig.~\ref{fig:experiment}). The experiments were running for one week each, with a typical beam current of 1~nA. 

The experimental setup enables particle-FF-$\gamma$ coincidences that, together with energy and time information, are sorted event-by-event. In the present work, we focused on the $^{233}$U(d,pf) and the $^{239}$Pu(d,pf) reactions. The detection of a charged particle in SiRi was the event trigger. In a timing interval of $\sim$20 ns we require a $\gamma$-signal in CACTUS and a FF in NIFF. 

From kinematics the measured energy of the outgoing, charged particle is converted into initial excitation energy, $E_\mathrm{x}$, of the fissioning system. In our cases, we measure the deposited energy of the proton in the particle telescope, thereby selecting $^{234}$U* and $^{240}$Pu* as the fissioning system, for experiments (A) and (B), respectively. The excitation energy was reconstructed event-by-event from the detected proton energy and emission angle, and accounting for energy losses in the target and backing. For each energy bin in $E_\mathrm{x}$, a correction for the neutron contribution to the $\gamma$-ray spectrum is performed, which is detailed in the next section. Finally, the raw $\gamma$-spectra are corrected for the detector response to produce a set of unfolded PFGS. The applied unfolding process, which has the advantage that the original, statistical fluctuations are preserved, is fully described in \citep{guttormsen1996unfolding}. NaI response functions are based on in-beam $\gamma$-lines from excited states in $^{56,57}$Fe, $^{28}$Si, $^{17}$O, and $^{13}$C, which were re-measured in 2012 \citep{PhysRevC.94.044321}.

\subsection{Correction for neutron contribution}
In the fission process, both neutrons and $\gamma$-rays are emitted. Neutrons can interact with the NaI crystals of CACTUS, depositing energy mostly in the form of $\gamma$-rays from (n,n$^{\prime}\gamma$) reactions. Unfortunately, the timing gate (20 ns) of the current set-up (Fig. \ref{fig:experiment}) only allows for discrimination between $\gamma$-rays and neutrons via time-of-flight (TOF) for the slowest neutrons, i.e. with energies lower than 600 keV. However, the majority of prompt neutrons emitted in fission have higher energy than this. To obtain PFGS, a correction for the neutron component needs to be made, with subtraction of counts arising from energy deposition by neutrons. 

Our neutron correction method relies on using a neutron response spectrum of a NaI detector, which is most representative of that for fission neutrons. Normalizing this to the known average neutron multiplicity emitted in fission for a particular compound nucleus excitation energy allows estimation of the neutron component in the total measured PFGS at this energy. This component is then subtracted. In this work we used a spectrum \citep{hausser1983prompt} for 2.3 MeV neutrons, which is close to the average fission neutron energy.

The response of 7.6 cm$\times$7.6 cm (3$^{\prime\prime}\times3^{\prime\prime}$) NaI detectors to incident neutrons at energies between 0.4 and 10 MeV has been measured by H\"{a}usser et al. \citep{hausser1983prompt}, using a TOF discrimination with quasi-monoenergetic neutrons produced in the $^{7}$Li(p,n) and $^{197}$Au(p,n) reactions. They find that the neutron response is dominated by (n,n$^{\prime}\gamma$) reactions. For the energies most prominent from fission neutrons, 1-2.5 MeV, most counts in the NaI detectors are observed between 0.4 and 1 MeV. For 2.3 MeV neutrons, they report 0.13(5) triggers per incident neutron. Since the CACTUS detectors are longer (12.5 cm), we scale the number of triggers to 0.21(8) triggers per incident neutron. We assume that the intrinsic detection efficiency, $\epsilon_{int}$, for $\gamma$-rays from fission is the same as those created in the detector by (n,n$^{\prime}\gamma$) reactions. The $\gamma$-ray multiplicity, $\bar{M}$, for neutron contribution correction purposes is taken as 6.31 for $^{234}$U* \citep{pleasonton1973prompt}  and 7.15 for $^{240}$Pu* \citep{ullmann2013prompt}.

\begin{table}
\caption{\label{tab:table3}%
Parameters to scale the excitation energy dependence of the average total neutron multiplicity relative to the neutron separation energy $S_\mathrm{n}$ extracted from Ref. \cite{ENDF233U} ($^{234}$U*) and Ref. \citep{ENDF239Pu} ( $^{240}$Pu*).}

\begin{ruledtabular}
\begin{tabular}{lcc}
 & $^{234}$U* & $^{240}$Pu*\\
\hline
a (n/MeV)& 0.1 &0.14 \\
b ($\bar\nu$ @ thermal fission) & 2.5 & 2.9\\
 $S_\mathrm{n}$ (MeV) & 6.85 & 6.53
\end{tabular}
\end{ruledtabular}
\end{table}

\begin{figure}
\centering
\includegraphics[width=0.5\textwidth]{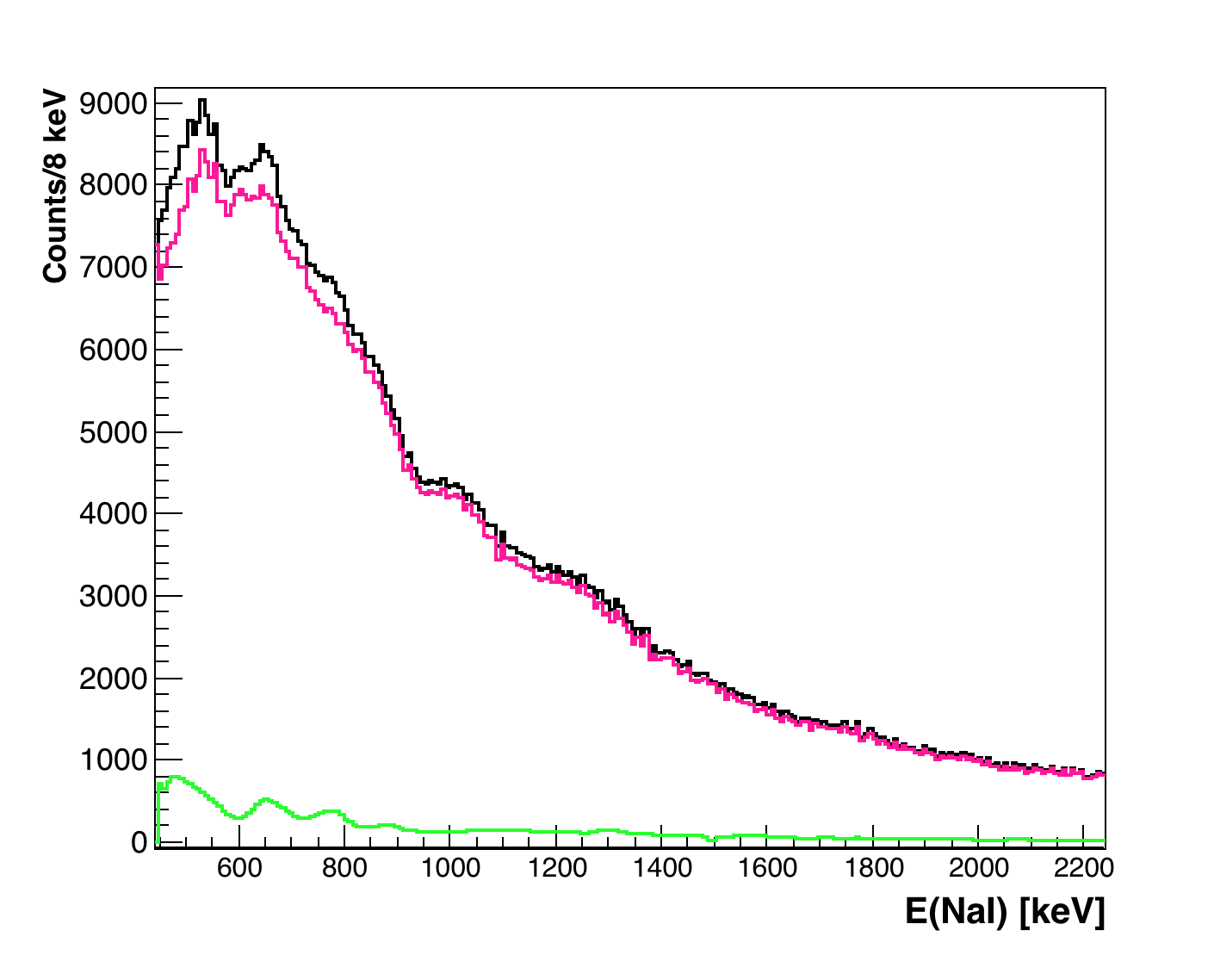}
\caption{(Color online) The total (summed over all $E_\mathrm{x}$) raw PFGS detected in the $^{233}$U(d,pf) reaction (black) and the calculated spectral contribution due to interactions of prompt fission neutrons in the NaI detector (green). The corrected $\gamma$-spectrum is also shown (pink).
\label{fig:neutron_correction}}
\end{figure}

The relative contribution, $f$, of neutrons to the measured data 
$N_{\mathrm{tot}}(E_\mathrm{x},E_\gamma)$ for each excitation energy 
$E_\mathrm{x}$ and $\gamma$-ray energy bin $E_\gamma$ can be 
estimated by the detection efficiencies. Taking into account the ratio 
of neutron and $\gamma$-ray multiplicities we find
\begin{align}
f = \frac {\epsilon_\mathrm{int,n}  
\bar\nu}{\epsilon_\mathrm{int,n} \bar\nu + \bar M}.
\end{align}
The neutron  multiplicity $\bar\nu$ is known to vary approximately linearly as a 
function of the incident neutron energy $E_\mathrm {n}$ \citep{madland2006total, ENDF233U, ENDF239Pu}. Taking into 
account the neutron separation energy $S_\mathrm{n}$, the same dependence 
is assumed for the compound nucleus excitation energy $E_\mathrm{x}$ 
with the parameters given in Table \ref{tab:table3}, such that 
$\bar\nu(E_\mathrm{x}) = a(E_\mathrm{x}-S_\mathrm{n}) +b$. The total contribution to the data caused by neutrons is estimated as a fraction of counts, $f(E_\text{x})$, that is weighted as a function of E$_{\gamma}$ by H\"{a}usser’s neutron spectrum $H(E_\gamma)$, i.e. 
\begin{align}
N_\mathrm{n}(E_\mathrm{x},E_\gamma)
= N_\mathrm{tot}(E_\mathrm{x}) f(E_\mathrm{x}) H(E_\gamma),
\end{align}
where $N_\text{tot}(E_\text{x})$ is the projection of the $\gamma$-matrix onto $E_\text{x}$
\begin{align}
N_\mathrm{tot}(E_\mathrm{x}) = 
\sum_{E_\gamma} N_\mathrm{tot}(E_\mathrm{x},E_\gamma).
\end{align}
{$N_\mathrm{tot}(E_\mathrm{x},E_\gamma)$ is the matrix element in the $\gamma$-matrix. $H(E_\gamma)$ is normalized so that $\sum_{E_\gamma}H(E_\gamma) = 1.$ The $\gamma$-ray spectrum $N_\gamma(E_\mathrm{x},E_\gamma)$ is obtained by subtracting the neutron contribution $N_\mathrm{n}(E_\mathrm{x},E_\gamma)$ from the measured data $N_\mathrm{tot}(E_\mathrm{x}, E_\gamma)$}
\begin{align}
N_\gamma(E_\mathrm{x},E_\gamma) = N_\mathrm{tot}(E_\mathrm{x},E_\gamma) 
-  N_\mathrm{n}(E_\mathrm{x},E_\gamma).
\end{align}
The results of the subtraction procedure can be seen graphically in Fig. \ref{fig:neutron_correction}, where the raw spectrum, neutron contribution and corrected spectrum are shown. Since inelastic scattering is the main energy deposition mechanism for neutrons, which occurs mostly on low-lying states in sodium and iodine nuclei, the neutron contribution is largest in the low energy part of the spectrum. However, overall, the correction for neutron contribution in our experiments remains small (see Table \ref{tab:table2}).

\begin{table}
\caption{\label{tab:table2}%
Values used for calculating the neutrons in the CACTUS detectors. The average neutron energies were calculated from ENDF/B VII.1 \citep{chadwick2011endf}. Neutron multiplicities $\bar\nu$ are taken from Ref. \citep{ENDF233U, ENDF239Pu} and $\gamma$-ray multiplicities $\bar{M}$, from Ref. \citep{pleasonton1973prompt} ($^{234}$U*) and Ref. \citep{ullmann2013prompt} ($^{240}$Pu*).}
\begin{ruledtabular}
\begin{tabular}{lcc}
 & A ($^{234}$U*) & B ($^{240}$Pu*)\\
\hline
Average neutron energy (MeV)& 2.0 & 2.1 \\[0.3em]

Intrinsic neutron efficiency & & \\ 
(triggers/neutron) & 0.21(8) & 0.21(8)\\[0.3em]

Neutron multiplicities & &\\ 
(@ thermal fission) & 2.5 & 2.9 \\[0.3em]

$\gamma$-ray multiplicities& 6.31(30) & 7.15(9) \\ [0.3em]

Relative contribution & &\\ 
(@ thermal fission) & 0.0768 & 0.078 \\ [0.3em]

\end{tabular}
\end{ruledtabular}
\end{table}

\subsection{Extrapolation of spectra towards zero energy \label{sec:extrapolation}}
Detectors used in experiments that attempt to measure PFGS will always have an energy threshold to prevent rapid triggering on noise. Below this threshold $\gamma$-ray detection is impossible, so the lowest energy $\gamma$-rays emitted in fission will not be detected. As a consequence, this will introduce a systematic uncertainty in the deduction of average spectral quantities: Measured multiplicities $\bar{M}$ and total $\gamma$-energy $E_\mathrm{tot}$ will thus be lower, and measured average $\gamma$-ray energy $E_\mathrm{av}$ released per fission will be higher, than their actual values. In fact, such systematic uncertainties from threshold effects may explain discrepancies between previous PFG experimental results, \citep{Oberstedt2013, chyzh2012evidence}.	To account for the undetected $\gamma$-rays below threshold, it is necessary to make an extrapolation towards zero energy, such as e.g. that performed in Ref. \citep{lebois2015comparative}. 
In our case the detection threshold was rather high, at 450 keV. As the shape of the $\gamma$ ray spectrum is not known for the low $\gamma$-ray energies, we  chose a constant value for the bins below threshold.
A reasonable extrapolation of each spectrum was made by averaging over the first three $\gamma$-ray bins above the threshold. The uncertainty was estimated by the minimum and maximum values in these bins, including their uncertainties. This results in an average value of about $5.5\pm2$ photons per fission per MeV ($^{234}$U*), below threshold. By assuming a non-zero value for this energy bin, the extrapolation reduces the uncertainty, but it does not eliminate it entirely. In our case it is still the dominant source of uncertainty on the absolute values of the average spectral quantities deduced. 
Since we compare our data with thermal neutron induced fission experiments, we chose the same cutoff of the PFGS as Ref. \citep{verbinski1973prompt}, of $E_\gamma=$140 keV.  

\section{Predictions with the GEF code}
We compare our data to predictions from the semi-empirical GEneral Fission model (GEF) \citep{schmidt2016general}. GEF is based on the observation of a number 
of regularities in fission observables, revealed by experimental studies, combined with general 
laws of statistical and quantum mechanics. It provides a general description of essentially all 
fission observables (fission-fragment yields and kinetic energies, prompt and delayed neutrons and
$\gamma$-rays, and isomeric ratios) in a consistent way while preserving the correlations between 
all of them. GEF has shown to be able to explain in an unprecendented good manner fission fragment
and neutron properties over a wide range, running from spontaneous fission to induced fission up to 
an excitation energy of about 100 MeV for Z = 80 to Z = 112 \citep{schmidt2016general}. Modelling of $\gamma$-rays in
fission has been implemented most recently. In contrast to other existing codes in the field,
GEF provides also reliable predictions for nuclei for which no experimental data exist. 
This is particularly important in our case, since no experimental data on the fragment properties exist 
for the majority of the excitation energies we are investigating. 

Calculations were performed for fission of both $^{234}$U* and $^{240}$Pu*, applying the same cutoff of the PFGS as for the experimental data, of 140 keV, as described in section \nolinebreak\ref{sec:extrapolation}. The total angular momentum $J=I_0 + L_\mathrm{trans}$ is the sum of the target nucleus ground state spin $I_0$ and the angular momentum $L_\mathrm{trans}$ transferred in the (d,p) reaction. The distribution in the GEF v.2016/1.1 calculations is given by 
\begin{align}
\rho(J) \propto (2 J  + 1) \exp(-J(J+1)/J^2_\mathrm{rms}),
\end{align}

where we used the root mean square (rms) of the total angular momentum\footnote{$J_\mathrm{rms}$ can be expressed in terms of the spin cut-off parameter $\sigma$ by $J_\mathrm{rms}=\sqrt{2}\sigma$} $J_\mathrm{rms}$ and the excitation energy to describe the fissioning system as input. The maximum value for $J_\mathrm{rms}$ of 12 was obtained from $J_\mathrm{rms}= \sqrt{2T\mathscr{I}}/\hbar$, \cite{Ericson1960} where the nuclear temperature was chosen to be $T \approx 0.45$ MeV in line with other actinide nuclei \citep{guttormsen2013constant,guttormsen2014scissors}. The rigid body of moment of inertia $\mathscr{I}$ is given by $\frac{2}{5} m_\mathrm{A} (r_0 A^{1/3})^2 (1+0.31 \beta_2) \approx 160 (\hbar c)^2/MeV$, where we used the isotope mass $m_\mathrm{A}$, the mass number $A$, the quadrupole deformation $\beta_2$ from Ref. \cite{capote2009ripl}, and radius parameter $r_0 \lesssim 1.3$. The results are compared to an intermediate value of $J_\mathrm{rms} = 8$, and to the lower limit, $J_\mathrm{rms} = 0$, where the latter facilitates the comparison to neutron induced reactions, which transfer little angular momentum. Additionally we performed calculations for a energy dependent $J_\mathrm{rms}$, which was adopted from the systematics of Ref. \cite{bucurescu2005systematics}
\begin{align}
J_\mathrm{rms}^2(E_\mathrm{x}) = 2 \times  0.0146 \, A^{5/3} \frac{  1 + \sqrt{1 + 4 a (E_\mathrm{x}-E_1)} }{2a},
\end{align}
where the level density parameter $a$ and the energy backshift $E_1$ are obtained from a fit to experimental data \cite{bucurescu2005systematics}.

\section{Experimental results}

\subsection{The $^{234}$U* case}
Fig. 3 shows a three-dimensional overview of the data set where, for a given compound nucleus excitation energy, the corresponding raw detected PFGS (prior to unfolding the response function) is displayed with the neutron contribution subtracted. The excitation energy range, over which the data are collected, can be seen more closely in Fig. 4., which histograms the double coincidences of protons and fission fragments (d,pf) and triple coincidences of  protons, fission fragments and $\gamma$-rays (d,pf$\gamma$) as a function of $E_\text{x}$. In the case of $^{234}$U*, only a very few sub-threshold fission events occur below the inner fission barrier \citep{capote2009ripl} at $E_\text{x}$ = 4.8 MeV, which is 2 MeV below the neutron separation energy at 6.85 MeV \citep{browne2007nuclear}. The $^{233}$U(d,pf) reaction at 12.5 MeV incident energy populates compound nuclear excitation energies up to a maximum of 10 MeV in this case.

\begin{figure}
\centering
\includegraphics[width=0.5\textwidth]{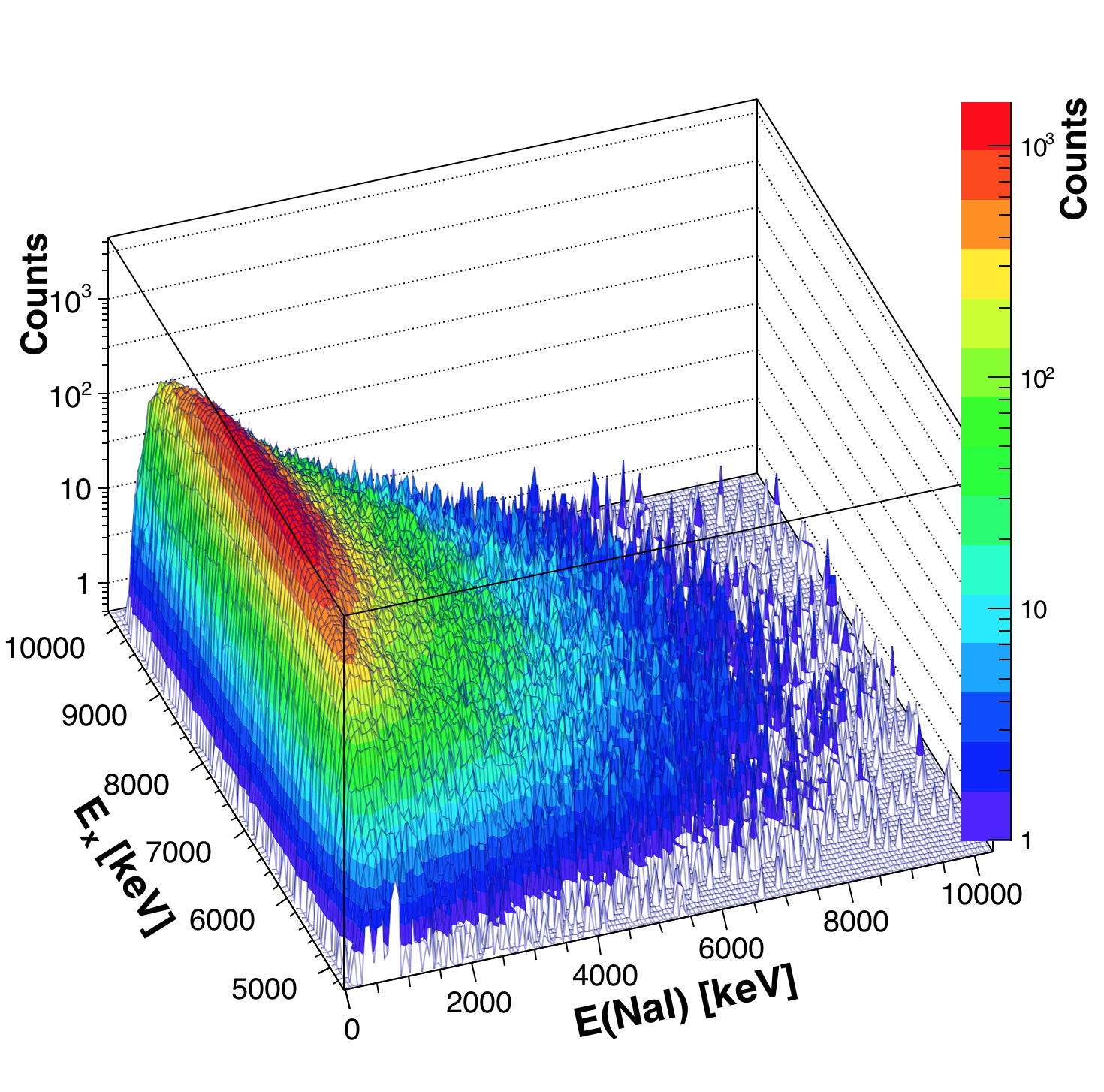}
\caption{(Color online) Matrix of the fission and proton gated raw $\gamma$-data from the $^{233}$U(d,pf) reaction (after subtraction of the contribution from neutrons). The x-axis gives the deduced compound nucleus excitation energy E$_{x}$. The y-axis gives the detected $\gamma$-ray energy, and the z-axis gives the number of counts recorded during the experiment (not efficiency corrected). The bin width is 64 keV for $E_\mathrm{x}$ and $E_\gamma$.}
\label{fig:raw_gamma}
\end{figure}

\begin{figure}
\centering
\includegraphics[width=0.49\textwidth]{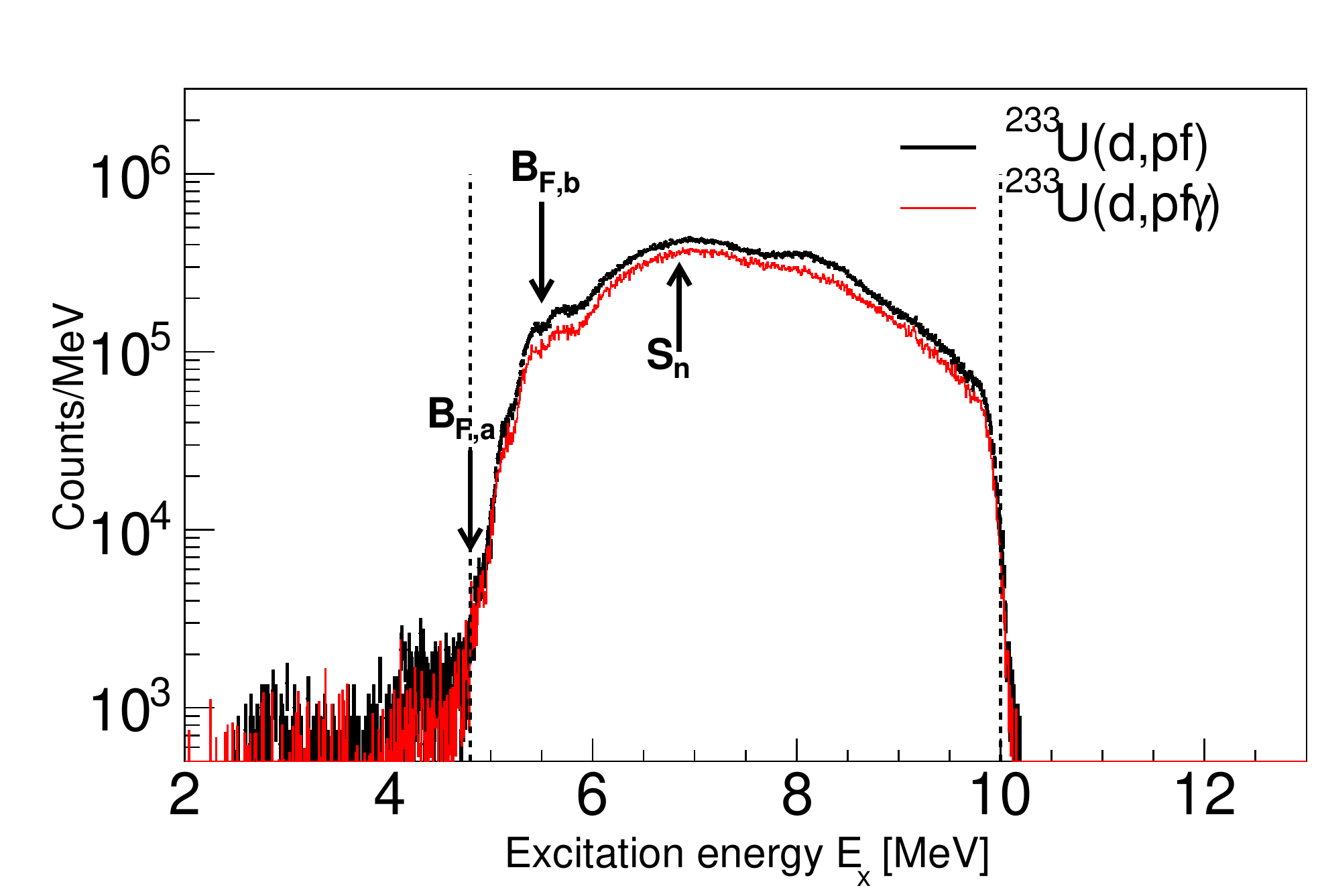}
\caption{(Color online) The total number of $^{233}$U(d,pf) and $^{233}$U(d,pf$\gamma$) events recorded during the experiment histogrammed as a function of the deduced compound nuclear excitation energy of $^{234}$U* for each event. The inner and outer fission barrier, $B_\mathrm{F,a}$ at 4.80 MeV and $B_\mathrm{F,b}$ at 5.50 MeV, and the neutron separation energy, S$_{n}$, at 6.85 MeV are shown. The dotted lines indicates the minimum and maximum $E_\mathrm{x}$ of the analysed area. The lower limit on $E_\mathrm{x}$ is the inner fission barrier.}
\label{fig:Histogram_U3}
\end{figure}

\begin{figure}
\centering
\includegraphics[width=0.5\textwidth]{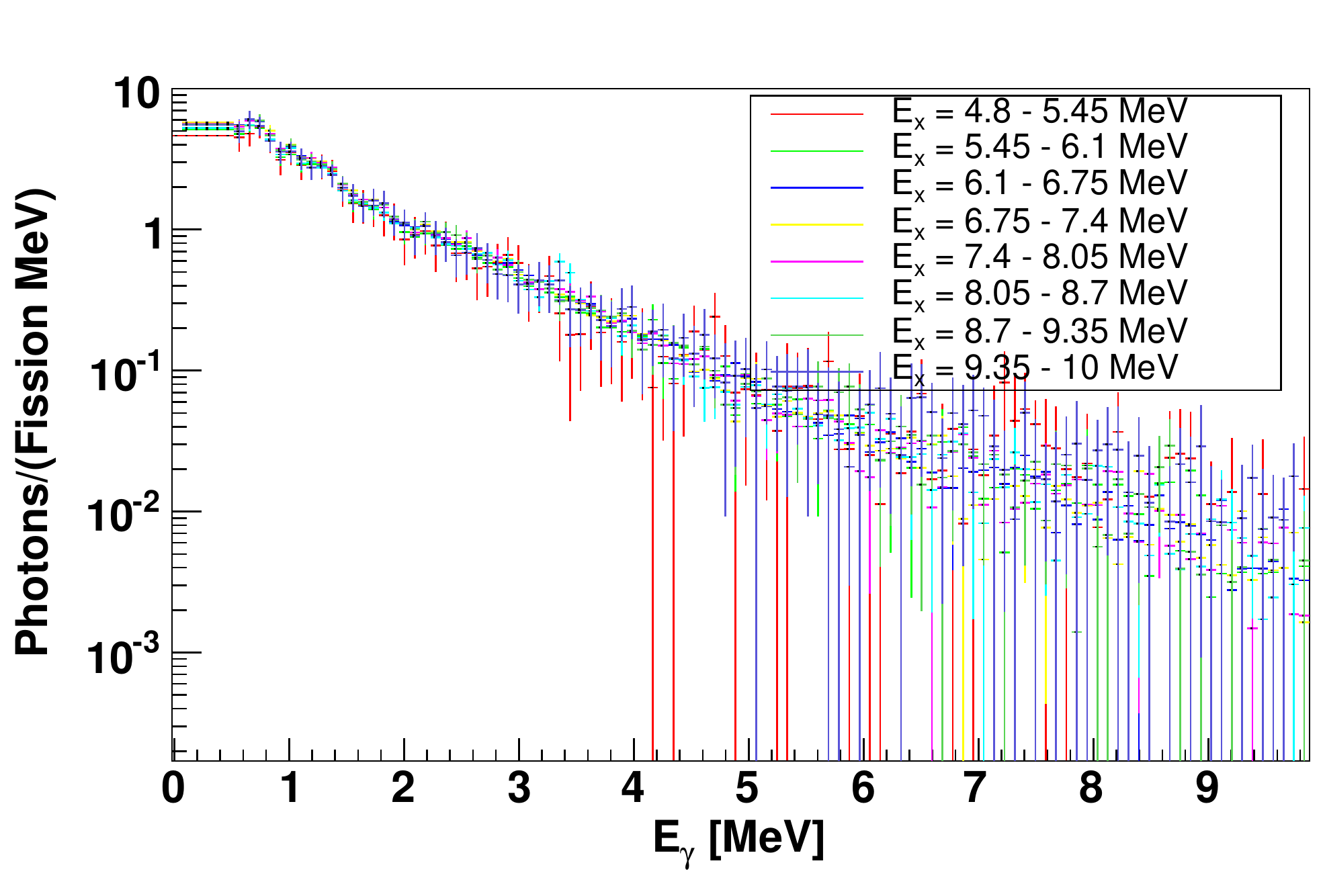}
\caption{(Color online) Overlay of the eight $^{233}$U(d,pf) PFGS for different excitation energy bins in compound nucleus excitation energy $E_\mathrm{x}$. The spectra are normalized to the number of photons per fission and per MeV to provide a comparison of the spectral shapes. The extrapolation from the detector threshold at 450 keV towards zero energy is explained in the text. 
}
\label{fig:PFGS_U3}
\end{figure}

The $E_\mathrm{x}$ range is divided into 8 bins, each with a width of 650 keV to obtain a sufficient statistics PFGS for each bin. Each spectrum is unfolded for the CACTUS response, and normalized to the number of fission events detected in that excitation energy bin. The set of eight normalized spectra is overlaid in Fig. \ref{fig:PFGS_U3}, and they exhibit similar spectral shapes.

\begin{figure}

\centering
\includegraphics[width=0.5\textwidth]{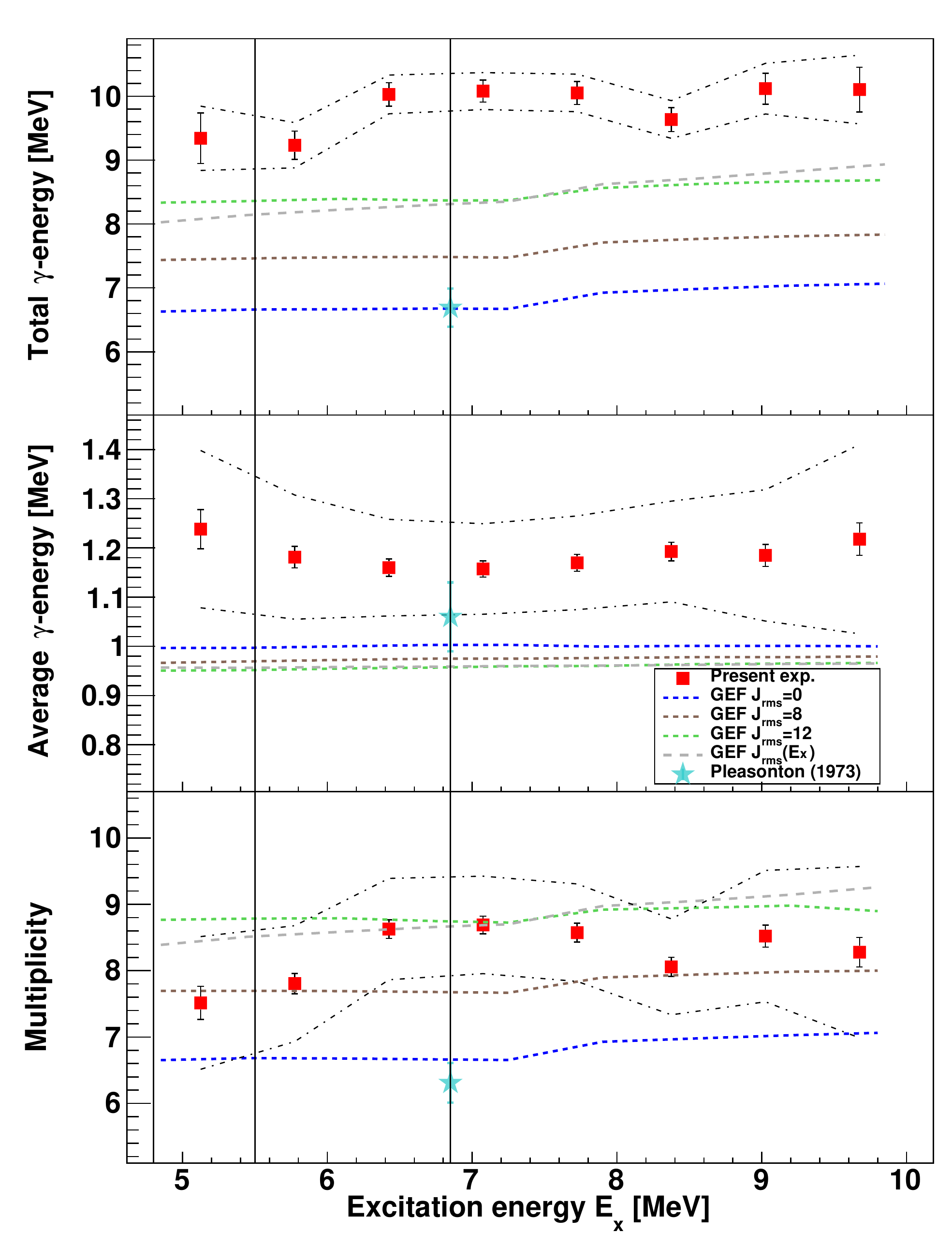}
\caption{(Color online) Energy dependence of the $^{233}$U(d,pf) average PFG spectral quantities compared with calculations from the GEF code for different $J_\mathrm{rms}$ of the $^{234}$U* nucleus. In addition, results from Pleasonton \citep{pleasonton1973prompt} are shown. Multiplicity, average $\gamma$-ray energy, and total $\gamma$-ray energy, as function of excitation energy of $^{234}$U* are shown. The error bars represent the statistical uncertainty of the measurement.  The dash-dotted lines represent the total uncertainty, which is the sum of the systematic uncertainty on the absolute values due to the detector threshold, and the extrapolation towards zero energy plus the statistical uncertainty. Vertical lines mark the inner and outer fission barriers ($E_\mathrm{x}$ = 4.8~MeV and $E_\mathrm{x}$ = 5.40~MeV) and the neutron separation energy ($E_\mathrm{x}$ = 6.85~MeV), respectively.}
\label{fig:Multiplicities_U3}
\end{figure}

The average spectral quantities after extrapolation to zero energy are then deduced and plotted as a function of the excitation energy. These results are plotted in Fig. \ref{fig:Multiplicities_U3} with their corresponding statistical error bars and compared with calculations from the GEF code. 
The wider band denoted by the dash-dotted lines indicates the sum of the statistical uncertainties on each data point plus the systematic uncertainty on the absolute values due to the presence of the detection threshold. 

\subsection{The $^{240}$Pu* case}
\begin{figure}
\centering
\includegraphics[width=0.5\textwidth]{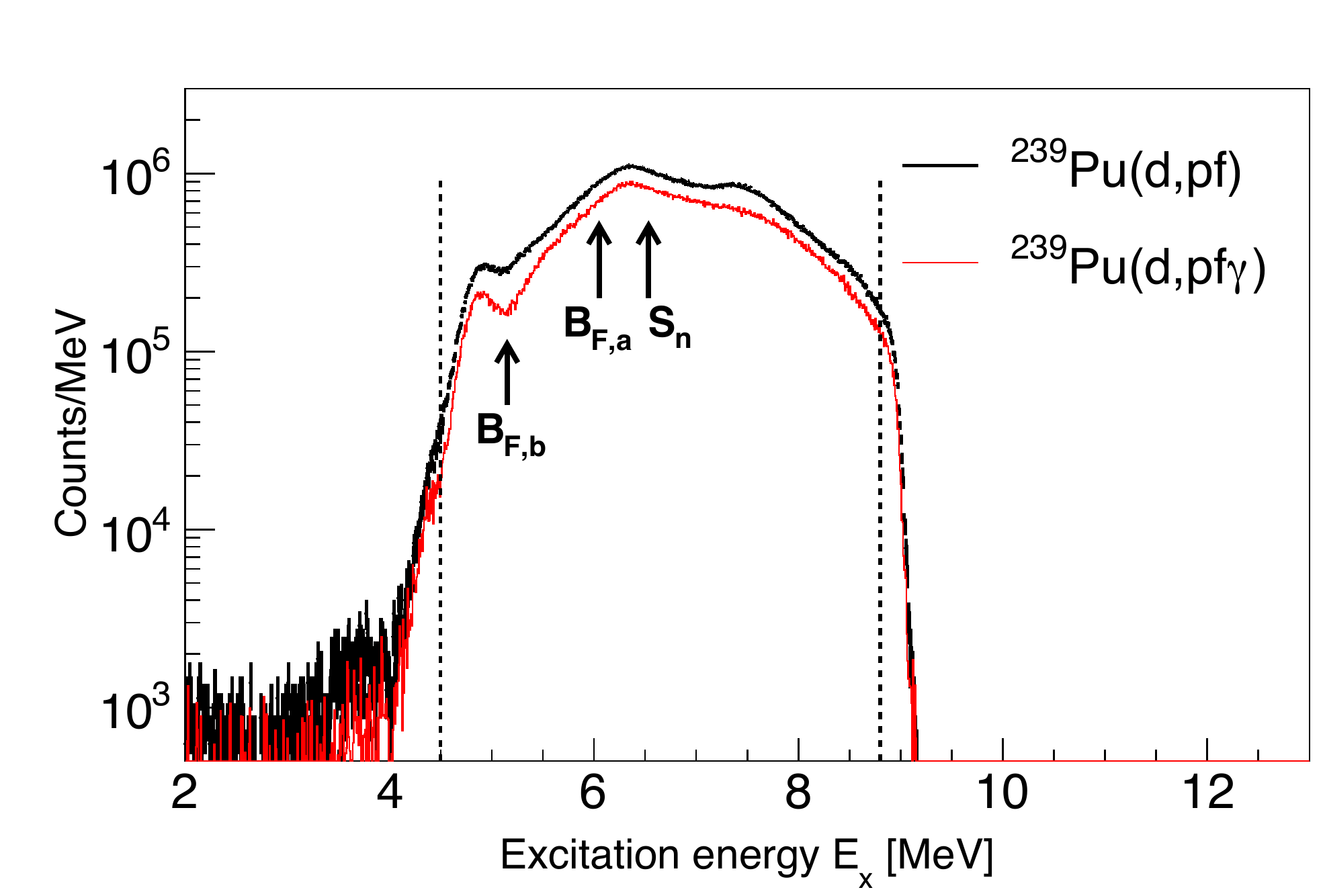}
\caption{(Color online) The total number of $^{239}$Pu(d,pf) and $^{239}$Pu(d,pf$\gamma$) events recorded during the experiment histogrammed as a function of the $^{240}$Pu* deduced excitation energy event-by-event. The inner and outer fission barrier, $B_\mathrm{F,a}$ at 6.05 MeV and $B_\mathrm{F,b}$ at 5.15 MeV, and the neutron separation energy, $S_\mathrm{n}$, at 6.53 MeV are shown. The dotted lines indicates the minimum and maximum $E_\mathrm{x}$ of the analysed area. The lower limit of $E_\mathrm{x}$ is at 4.8 MeV, which is more than 1 MeV below the fission barrier – due to sub barrier fission.}
\label{fig:Histogram_Pu9}
\end{figure}

The same analysis was performed for the $^{239}$Pu(d,pf) reaction. The (d,pf) and the (d,pf$\gamma$) reactions are histogrammed as functions excitation energy (Fig. \ref{fig:Histogram_Pu9}). In the $^{240}$Pu* case, there appears to be a significant amount of sub-barrier fission, which is in accordance with observations in Refs. \citep{hunyadi2001excited, glassel1976intermediate}. This can be explained in the double-humped fission barrier picture; by the resonant population of states in the second potential minimum of the $^{240}$Pu* nucleus and a tunnelling through the outer fission barrier.

\begin{figure}
\centering
\includegraphics[width=0.5\textwidth]{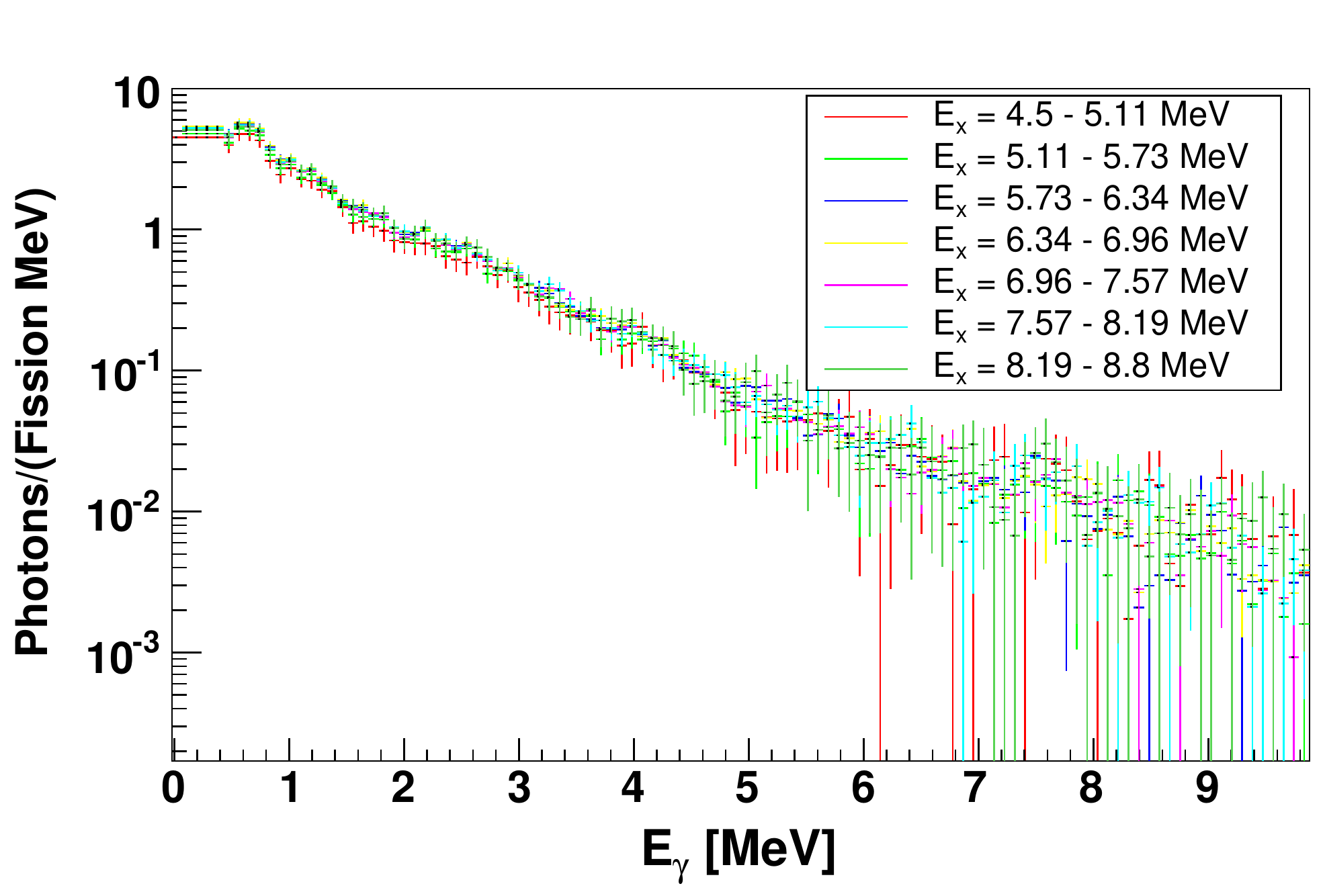}
\caption{(Color online) Overlay of the six $^{239}$Pu(d,pf) unfolded PFG gamma spectra for different excitation energy bins in compound nucleus excitation energy E$_{x}$. The spectra are normalized to the number of photons per fission and per MeV to provide a comparison of the spectral shapes. The extrapolation between 140 keV energy and the detector threshold at 450 keV is explained in the text. 
}
\label{fig:PFGS_Pu9}
\end{figure}

\begin{figure}
\centering
\includegraphics[width=0.5\textwidth]{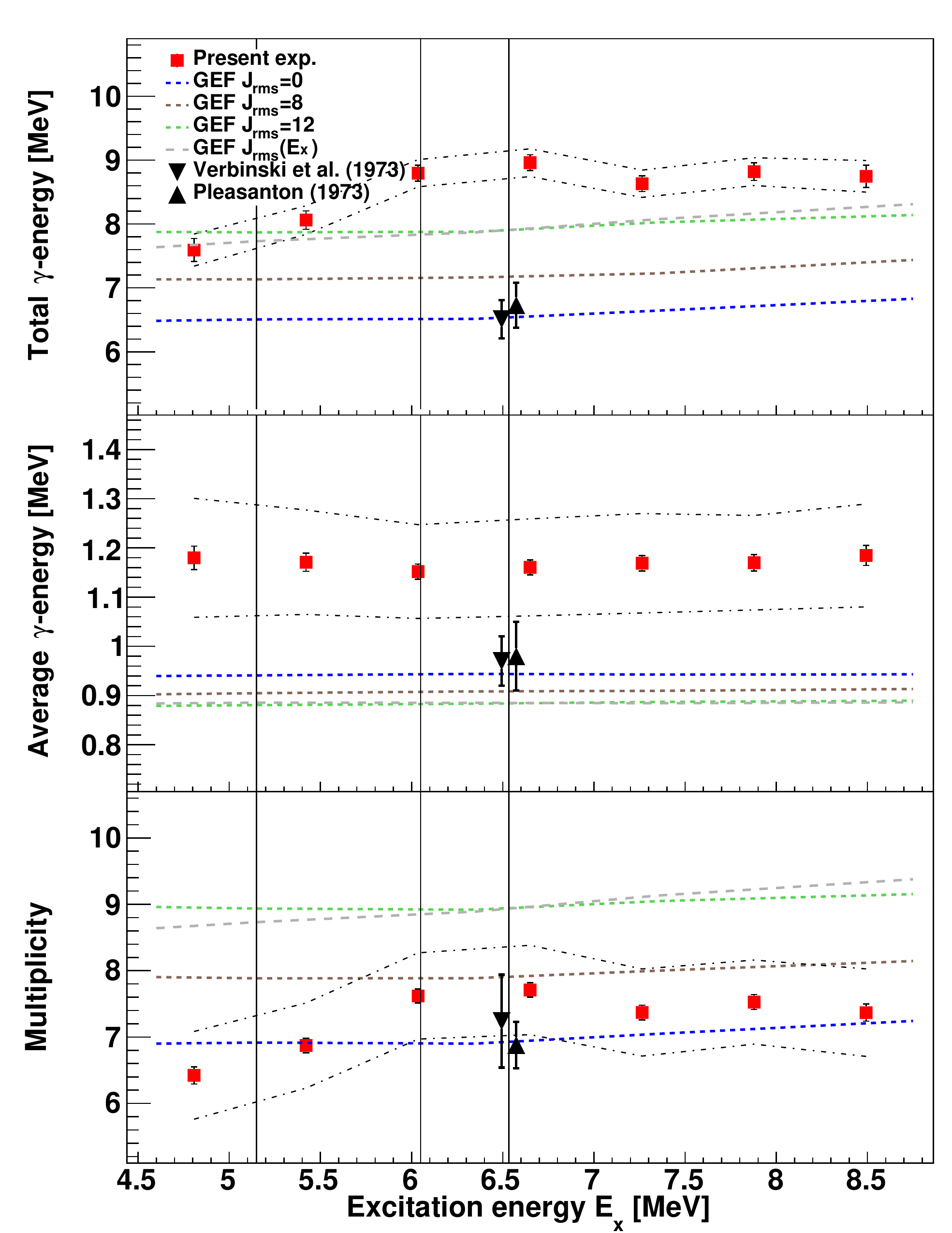}
\caption{(Color online) Energy dependence of the $^{239}$Pu(d,pf) PFG average spectral quantities from the GEF code for different $J_\mathrm{rms}$ of the $^{240}$Pu* nucleus. The thermal neutron data of Pleasonton (1973) \citep{pleasonton1973prompt} and Verbinski et al. (1973) \citep{verbinski1973prompt} are shift slightly around $S_\mathrm{n}$ for better visibility. Multiplicity, average $\gamma$-ray energy, and total $\gamma$-ray energy, as function of excitation energy of $^{240}$Pu* are shown. The error bars represent the statistical uncertainty of the measurement.  The dash-dotted lines represent the systematic uncertainty on the absolute values due to the detector threshold and the necessary extrapolation to zero energy. Vertical lines mark the inner and outer fission barriers ($E_\mathrm{x}$ = 6.05~MeV and $E_\mathrm{x}$ = 5.15~MeV) and the neutron separation energy ($E_\mathrm{x}$ = 6.5~MeV), respectively. }
\label{fig:Multiplicities_Pu9}
\end{figure}

The overlay of the unfolded PFGS for the $^{239}$Pu(d,pf) reaction is shown in Fig. \ref{fig:PFGS_Pu9}. The spectral shapes are all observed to be similar. However, the PFGS for the two lowest compound nucleus excitation energy bins starting at 4.65 MeV and 5.45 MeV appear to be significantly lower than the others. This effect also manifests itself in the average photon multiplicity $\bar M$ and total energy $E_\mathrm{tot}$ release at this energy (see Fig. \ref{fig:Multiplicities_Pu9}). We note that this is the region below the fission barrier and, hence, originates from sub-barrier fission. Otherwise, the trends for the spectral characteristics seem to have no significant trend and are fairly constant, i.e. independent of excitation energy and thus consistent with the predictions of the GEF code. 

\begin{figure}
\centering
\includegraphics[width=0.5\textwidth]{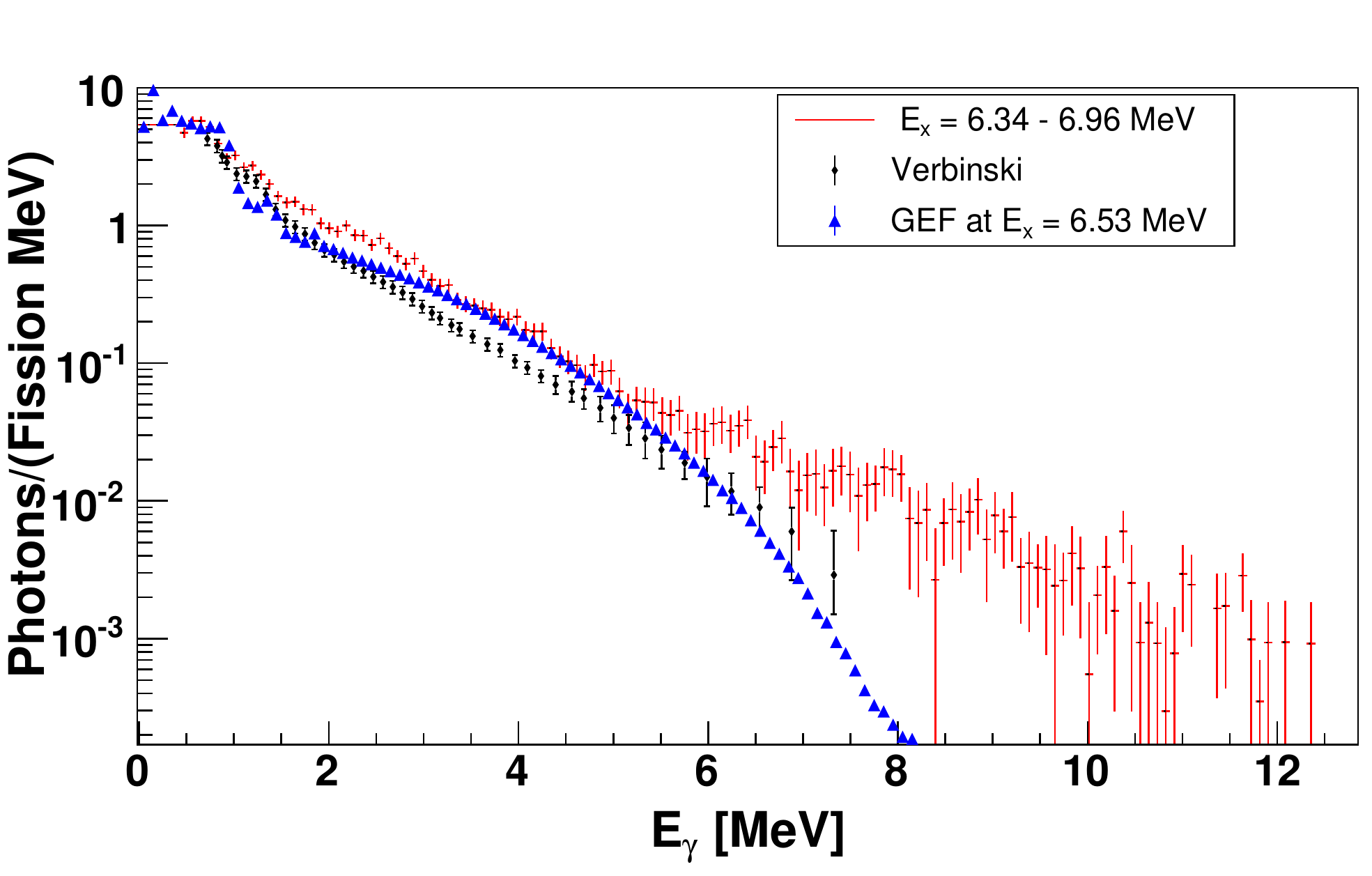}
\caption{(Color online) A comparison of the $^{239}$Pu(d,pf) PFGS measured at $E_\mathrm{x}\sim S_{n}$ (red), the PFGS for thermal neutron induced fission $^{239}$Pu(n$_\mathrm{th}$,f) from Verbinski  et al.\citep{verbinski1973prompt} (black points), and the calculations by GEF for $J_\mathrm{rms}=8$ and $E_\mathrm{x} = 6.35$(blue).
}
\label{fig:PFGS_Compare_Pu9}
\end{figure}

Finally, we compare the measured PFGS at excitation energy of 6.5 MeV, which corresponds to the thermal neutron induced fission reaction for $^{239}$Pu, with the measured PFGS of Verbinski et al. \citep{verbinski1973prompt} for thermal neutron induced fission. Fig. \ref{fig:PFGS_Compare_Pu9} shows this comparison along with a spectrum from the GEF code. An excess of counts is observed between 2 and 4 MeV for our “surrogate” PFGS measured in the $^{239}$Pu(d,pf) reaction as compared to the neutron induced reaction. 

\section{Discussion}
In this study, both experiments reveal an approximately constant behaviour of average $\gamma$-ray energy $E_\mathrm{av}$, $\bar{M}$, and $E_\mathrm{tot}$, as a function of $E_\mathrm{x}$ of the fissioning system; shown in Fig. 6 for uranium and Fig. 9 for plutonium. The constant trend (though not the absolute value) in spectral characteristics that we observe is broadly in line with the predictions with GEF.

There seems to be a slight decrease in the $\bar{M}$ below $S_\mathrm{n}$ for both nuclei, but more clearly seen in the plutonium data.
Although up to 5 MeV of extra excitation energy for the hot fission fragments is available, this energy is clearly more efficiently dissipated by the evaporation of prompt fission neutrons. The prompt fission neutron multiplicity is well known to increase linearly with excitation energy.
One could expect that the total angular momentum $J$ of the fissioning nucleus should increase with increasing $E_\mathrm{x}$.
Our experimental data exhibit a flat trend, which is compatible to GEF calculations for a constant or energy dependent $J_\mathrm{rms}$ in the studied excitation energy range.

An excess of counts is observed when comparing the surrogate (d,p) PFG and thermal neutron induced PFGS. Such a discrepancy might arise from differences in the surrogate and neutron induced reactions.
The spectrum (Fig. \ref{fig:PFGS_Compare_Pu9}) predicted with the GEF code lies in between the two experimental cases in the region in which the deviation is observed. For $\gamma$-rays above 8 MeV, significantly less photons are predicted in comparison with our data. The spectrum by Verbinski et al. \citep{verbinski1973prompt} is reported only up to $E_{\gamma}$=7.5 MeV.

It is expected that reactions involving charged particles will on average introduce more angular momentum $L_\mathrm{trans}$ into the reaction, than thermal neutron induced reactions. The distribution of the angular momentum $J$ will have a tail, which extends higher, the greater the mass difference is between the ingoing and outgoing charged particles in the reaction. It may, therefore, be possible that the excess counts observed in the PFGS of the “surrogate” reaction is an angular momentum effect introduced by using the (d,p) reaction to induce fission, instead of neutrons. 

It is consistent that for $\bar{M}$ and $E_\mathrm{tot}$ our (d,p) PFG data are in better agreement with larger $J_\mathrm{rms}$, whereas the thermal neutron induced data are in all cases in good agreement with low $J_\mathrm{rms}$. For $E_\mathrm{av}$ the results of the GEF calculation are in both reactions less sensitive to $J_\mathrm{rms}$, and there discrepancy between our experimental results and the calculations increases.

The absolute values of the $E_\mathrm{tot}$ and the $\bar{M}$ are higher for the $^{234}$U* than the $^{240}$Pu*. Comparison with the results from GEF, and a slightly higher deuteron beam energy, indicates a higher angular momentum in the uranium case.
Average higher angular momentum of the fission fragments might result in neutron emission being partially hindered from odd fission fragments up to 1 MeV above their $S_\mathrm{n}$. In such a case $\gamma$-ray emission will compete with neutron emission, also above $S_\mathrm{n}$. This would result in an increased total $\gamma$ energy and higher $\bar{M}$. 

Recently, surrogate measurements have demonstrated that radiative 
capture and fission cross sections \citep{boutoux2012study} can be 
used to get quantitative insight
into the angular momentum $L_\mathrm{trans}$ imparted to the compound nucleus 
following a specific transfer reaction. A detailed review of both theory, experimental results and challenges can be found in Ref. \citep{escher2012compound}.
The connection between these cross sections and $L_\mathrm{trans}$ involves 
sophisticated Hauser-Feshbach calculations \citep{hauser1952inelastic}.
On the other hand, it is established that prompt fission $\gamma$ 
multiplicity $\bar{M}$ is the most direct probe of the angular momentum of the fission fragments. The latter is influenced by the angular momentum of
the fissioning system, i.e. $L_\mathrm{trans}$ in the presented GEF calculations. The present
work shows for the first time that the measured $\bar{M}$ is indeed 
sensitive to $L_\mathrm{trans}$. Hence, it can be
used as an alternative observable, complementary to cross sections \citep{boutoux2012study}, to 
quantify $L_\mathrm{trans}$.

Above the neutron binding energy $S_\mathrm{n}$ there is no significant increase in average PFG energy and total PFG energy released per fission with increasing excitation $E_\mathrm{x}$. This observation is important for applications, since $\gamma$-rays from fission are responsible for a large part of the heating that occurs in reactor cores. The observed result implies that passing from current Generation-III thermal reactors to fast Generation-IV reactor concepts will not require significant changes in the modelling of $\gamma$ heat transport from the fast neutron induced fission process. Since $^{233}$U is the main fissile isotope in the thorium cycle, and $^{239}$Pu is the main fissile isotope in the plutonium/uranium cycle, and the flat trend is observed in both these nuclei, effects of $\gamma$ heating from fission in both cycles are expected to be similar.

\section{Conclusion}
Emission of prompt $\gamma$-rays from nuclear fission induced via the $^{233}$U(d,pf) and $^{239}$Pu(d,pf) reactions have been studied. PFGS have been extracted as functions of the compound nucleus excitation energy for both nuclei. The average spectral characteristics have been deduced and trends as a function of excitation energy have been studied and compared with calculations by the GEF code. 

We observe an approximately constant behaviour of the spectral properties as a function of energy for both nuclei. However, a much lower multiplicity is seen in the sub-barrier fission of $^{240}$Pu*. More detailed studies are needed to understand why sub-barrier fission results in emission of low multiplicities of prompt $\gamma$-rays from the excited fission fragments.  Furthermore, we observe an excess of $\gamma$-rays above 2 MeV emitted in the surrogate $^{239}$Pu(d,pf) reaction  when comparing to the neutron induced PFGS measured by Verbinski et al. This effect is not yet understood, but may be as due to higher angular momenta involved in the transfer-induced reactions as compared to the neutron-induced one, over the energy range of our study. This conjecture is supported by GEF calculations.

Our measured $\gamma$ ray multiplicities and total $\gamma$ energies are higher than those observed for the neutron induced reactions from Verbinski et al. and Pleasonton. This difference may be explained as due to higher $J$ by comparing to the GEF calculations.

In the future we hope to revisit these types of measurements with the OSCAR array of 26 large volume LaBr$_{3}$ detectors currently being constructed at the Oslo Cyclotron Laboratory. These will not only provide a much better $\gamma$-ray energy resolution and lower energy thresholds, but an excellent timing resolution which will allow for discrimination of neutrons from $\gamma$-rays via time of flight. 

\begin{acknowledgments}
We would like to thank J. M{\"u}ller, E. A. Olsen, and A. Semchenkov for providing excellent beams during the experiments.We would also like to thank Beatriz Jurado and Karl-Heinz Schmidt for fruitful discussions.
A.C. Larsen gratefully acknowledges funding through ERC-STG-2014, grant agreement no. 637686. G.M.T. and S. Siem gratefully acknowledge funding through the Research Council of Norway, Project No. 222287 and grant no. 210007, respectively.
We would like to thank Lawrence Livermore National Laboratory for providing the $^{233}$U and $^{239}$Pu targets.
This work was performed under the auspices of the University of California Office of the President Laboratory Fees Research Program under Award No. 12-LR-238745, the U.S. Department of Energy by Lawrence Livermore National Laboratory under Contract DE-AC52-07NA27344, Lawrence Berkeley National Laboratory under Contract No. DE-AC02-05CH11231, and the National Research Foundation of South Africa under Grant Nos. 92789 and 83671.
M.Wiedeking acknowledges funding from the National Research Foundation of South Africa under Grant Nos. 92989 and 83867.

\end{acknowledgments}

\bibliography{references.bib}

\begin{thebibliography}{47}%
\makeatletter
\providecommand \@ifxundefined [1]{%
 \@ifx{#1\undefined}
}%
\providecommand \@ifnum [1]{%
 \ifnum #1\expandafter \@firstoftwo
 \else \expandafter \@secondoftwo
 \fi
}%
\providecommand \@ifx [1]{%
 \ifx #1\expandafter \@firstoftwo
 \else \expandafter \@secondoftwo
 \fi
}%
\providecommand \natexlab [1]{#1}%
\providecommand \enquote  [1]{``#1''}%
\providecommand \bibnamefont  [1]{#1}%
\providecommand \bibfnamefont [1]{#1}%
\providecommand \citenamefont [1]{#1}%
\providecommand \href@noop [0]{\@secondoftwo}%
\providecommand \href [0]{\begingroup \@sanitize@url \@href}%
\providecommand \@href[1]{\@@startlink{#1}\@@href}%
\providecommand \@@href[1]{\endgroup#1\@@endlink}%
\providecommand \@sanitize@url [0]{\catcode `\\12\catcode `\$12\catcode
  `\&12\catcode `\#12\catcode `\^12\catcode `\_12\catcode `\%12\relax}%
\providecommand \@@startlink[1]{}%
\providecommand \@@endlink[0]{}%
\providecommand \url  [0]{\begingroup\@sanitize@url \@url }%
\providecommand \@url [1]{\endgroup\@href {#1}{\urlprefix }}%
\providecommand \urlprefix  [0]{URL }%
\providecommand \Eprint [0]{\href }%
\providecommand \doibase [0]{http://dx.doi.org/}%
\providecommand \selectlanguage [0]{\@gobble}%
\providecommand \bibinfo  [0]{\@secondoftwo}%
\providecommand \bibfield  [0]{\@secondoftwo}%
\providecommand \translation [1]{[#1]}%
\providecommand \BibitemOpen [0]{}%
\providecommand \bibitemStop [0]{}%
\providecommand \bibitemNoStop [0]{.\EOS\space}%
\providecommand \EOS [0]{\spacefactor3000\relax}%
\providecommand \BibitemShut  [1]{\csname bibitem#1\endcsname}%
\let\auto@bib@innerbib\@empty
%</preamble>
\bibitem [{\citenamefont {Meitner}\ and\ \citenamefont
  {Frisch}(1939)}]{Meitner1939}%
  \BibitemOpen
  \bibfield  {author} {\bibinfo {author} {\bibfnamefont {L.}~\bibnamefont
  {Meitner}}\ and\ \bibinfo {author} {\bibfnamefont {O.~R.}\ \bibnamefont
  {Frisch}},\ }\href {\doibase 10.1038/143239a0} {\bibfield  {journal}
  {\bibinfo  {journal} {Nature}\ }\textbf {\bibinfo {volume} {143}},\ \bibinfo
  {pages} {239–240} (\bibinfo {year} {1939})}\BibitemShut {NoStop}%
\bibitem [{\citenamefont {Hahn}\ and\ \citenamefont
  {Strassmann}(1939)}]{Hahn1939}%
  \BibitemOpen
  \bibfield  {author} {\bibinfo {author} {\bibfnamefont {O.}~\bibnamefont
  {Hahn}}\ and\ \bibinfo {author} {\bibfnamefont {F.}~\bibnamefont
  {Strassmann}},\ }\href@noop {} {\bibfield  {journal} {\bibinfo  {journal}
  {Die Naturwissenschaften}\ }\textbf {\bibinfo {volume} {27}},\ \bibinfo
  {pages} {11} (\bibinfo {year} {1939})}\BibitemShut {NoStop}%
\bibitem [{\citenamefont {Bohr}\ and\ \citenamefont
  {Wheeler}(1939)}]{Bohr1939}%
  \BibitemOpen
  \bibfield  {author} {\bibinfo {author} {\bibfnamefont {N.}~\bibnamefont
  {Bohr}}\ and\ \bibinfo {author} {\bibfnamefont {J.~A.}\ \bibnamefont
  {Wheeler}},\ }\href {\doibase 10.1103/physrev.56.426} {\bibfield  {journal}
  {\bibinfo  {journal} {Physical Review}\ }\textbf {\bibinfo {volume} {56}},\
  \bibinfo {pages} {426–450} (\bibinfo {year} {1939})}\BibitemShut {NoStop}%
\bibitem [{\citenamefont {Pleasonton}\ \emph {et~al.}(1972)\citenamefont
  {Pleasonton}, \citenamefont {Ferguson},\ and\ \citenamefont
  {Schmitt}}]{pleasonton1972prompt}%
  \BibitemOpen
  \bibfield  {author} {\bibinfo {author} {\bibfnamefont {F.}~\bibnamefont
  {Pleasonton}}, \bibinfo {author} {\bibfnamefont {R.~L.}\ \bibnamefont
  {Ferguson}}, \ and\ \bibinfo {author} {\bibfnamefont {H.}~\bibnamefont
  {Schmitt}},\ }\href@noop {} {\bibfield  {journal} {\bibinfo  {journal}
  {Physical Review C}\ }\textbf {\bibinfo {volume} {6}},\ \bibinfo {pages}
  {1023} (\bibinfo {year} {1972})}\BibitemShut {NoStop}%
\bibitem [{\citenamefont {Bowman}\ and\ \citenamefont
  {Thompson}(1958)}]{bowman1958prompt}%
  \BibitemOpen
  \bibfield  {author} {\bibinfo {author} {\bibfnamefont {H.}~\bibnamefont
  {Bowman}}\ and\ \bibinfo {author} {\bibfnamefont {S.}~\bibnamefont
  {Thompson}},\ }\href@noop {} {\emph {\bibinfo {title} {The prompt radiations
  in the spontaneouos fission of $^{252}$Cf}}},\ \bibinfo {type} {Tech. Rep.}\
  (\bibinfo  {institution} {California. Univ., Livermore. Radiation Lab.},\
  \bibinfo {year} {1958})\BibitemShut {NoStop}%
\bibitem [{\citenamefont {Albinsson}(1971)}]{albinsson1971decay}%
  \BibitemOpen
  \bibfield  {author} {\bibinfo {author} {\bibfnamefont {H.}~\bibnamefont
  {Albinsson}},\ }\href@noop {} {\bibfield  {journal} {\bibinfo  {journal}
  {Physica Scripta}\ }\textbf {\bibinfo {volume} {3}},\ \bibinfo {pages} {113}
  (\bibinfo {year} {1971})}\BibitemShut {NoStop}%
\bibitem [{\citenamefont {Talou}\ \emph {et~al.}(2016)\citenamefont {Talou},
  \citenamefont {Kawano}, \citenamefont {Stetcu}, \citenamefont {Lestone},
  \citenamefont {McKigney},\ and\ \citenamefont {Chadwick}}]{Talou2016}%
  \BibitemOpen
  \bibfield  {author} {\bibinfo {author} {\bibfnamefont {P.}~\bibnamefont
  {Talou}}, \bibinfo {author} {\bibfnamefont {T.}~\bibnamefont {Kawano}},
  \bibinfo {author} {\bibfnamefont {I.}~\bibnamefont {Stetcu}}, \bibinfo
  {author} {\bibfnamefont {J.~P.}\ \bibnamefont {Lestone}}, \bibinfo {author}
  {\bibfnamefont {E.}~\bibnamefont {McKigney}}, \ and\ \bibinfo {author}
  {\bibfnamefont {M.~B.}\ \bibnamefont {Chadwick}},\ }\href {\doibase
  10.1103/PhysRevC.94.064613} {\bibfield  {journal} {\bibinfo  {journal} {Phys.
  Rev. C}\ }\textbf {\bibinfo {volume} {94}},\ \bibinfo {pages} {064613}
  (\bibinfo {year} {2016})}\BibitemShut {NoStop}%
\bibitem [{\citenamefont {Stetcu}\ \emph {et~al.}(2014)\citenamefont {Stetcu},
  \citenamefont {Talou}, \citenamefont {Kawano},\ and\ \citenamefont
  {Jandel}}]{stetcu2014properties}%
  \BibitemOpen
  \bibfield  {author} {\bibinfo {author} {\bibfnamefont {I.}~\bibnamefont
  {Stetcu}}, \bibinfo {author} {\bibfnamefont {P.}~\bibnamefont {Talou}},
  \bibinfo {author} {\bibfnamefont {T.}~\bibnamefont {Kawano}}, \ and\ \bibinfo
  {author} {\bibfnamefont {M.}~\bibnamefont {Jandel}},\ }\href@noop {}
  {\bibfield  {journal} {\bibinfo  {journal} {Physical Review C}\ }\textbf
  {\bibinfo {volume} {90}},\ \bibinfo {pages} {024617} (\bibinfo {year}
  {2014})}\BibitemShut {NoStop}%
\bibitem [{\citenamefont {Schmitt}\ \emph {et~al.}(1984)\citenamefont
  {Schmitt}, \citenamefont {Mouchaty},\ and\ \citenamefont
  {Haenni}}]{schmitt1984angular}%
  \BibitemOpen
  \bibfield  {author} {\bibinfo {author} {\bibfnamefont {R.}~\bibnamefont
  {Schmitt}}, \bibinfo {author} {\bibfnamefont {G.}~\bibnamefont {Mouchaty}}, \
  and\ \bibinfo {author} {\bibfnamefont {D.}~\bibnamefont {Haenni}},\
  }\href@noop {} {\bibfield  {journal} {\bibinfo  {journal} {Nuclear Physics
  A}\ }\textbf {\bibinfo {volume} {427}},\ \bibinfo {pages} {614} (\bibinfo
  {year} {1984})}\BibitemShut {NoStop}%
\bibitem [{\citenamefont {Oberstedt}\ \emph {et~al.}(2013)\citenamefont
  {Oberstedt}, \citenamefont {Belgya}, \citenamefont {Billnert}, \citenamefont
  {Borcea}, \citenamefont {Bry\ifmmode~\acute{s}\else \'{s}\fi{}},
  \citenamefont {Geerts}, \citenamefont {G\"o\"ok}, \citenamefont {Hambsch},
  \citenamefont {Kis}, \citenamefont {Martinez}, \citenamefont {Oberstedt},
  \citenamefont {Szentmiklosi}, \citenamefont {Tak\`acs},\ and\ \citenamefont
  {Vidali}}]{Oberstedt2013}%
  \BibitemOpen
  \bibfield  {author} {\bibinfo {author} {\bibfnamefont {A.}~\bibnamefont
  {Oberstedt}}, \bibinfo {author} {\bibfnamefont {T.}~\bibnamefont {Belgya}},
  \bibinfo {author} {\bibfnamefont {R.}~\bibnamefont {Billnert}}, \bibinfo
  {author} {\bibfnamefont {R.}~\bibnamefont {Borcea}}, \bibinfo {author}
  {\bibfnamefont {T.}~\bibnamefont {Bry\ifmmode~\acute{s}\else \'{s}\fi{}}},
  \bibinfo {author} {\bibfnamefont {W.}~\bibnamefont {Geerts}}, \bibinfo
  {author} {\bibfnamefont {A.}~\bibnamefont {G\"o\"ok}}, \bibinfo {author}
  {\bibfnamefont {F.-J.}\ \bibnamefont {Hambsch}}, \bibinfo {author}
  {\bibfnamefont {Z.}~\bibnamefont {Kis}}, \bibinfo {author} {\bibfnamefont
  {T.}~\bibnamefont {Martinez}}, \bibinfo {author} {\bibfnamefont
  {S.}~\bibnamefont {Oberstedt}}, \bibinfo {author} {\bibfnamefont
  {L.}~\bibnamefont {Szentmiklosi}}, \bibinfo {author} {\bibfnamefont
  {K.}~\bibnamefont {Tak\`acs}}, \ and\ \bibinfo {author} {\bibfnamefont
  {M.}~\bibnamefont {Vidali}},\ }\href {\doibase 10.1103/PhysRevC.87.051602}
  {\bibfield  {journal} {\bibinfo  {journal} {Phys. Rev. C}\ }\textbf {\bibinfo
  {volume} {87}},\ \bibinfo {pages} {051602} (\bibinfo {year}
  {2013})}\BibitemShut {NoStop}%
\bibitem [{\citenamefont {Chyzh}\ \emph {et~al.}(2013)\citenamefont {Chyzh},
  \citenamefont {Wu}, \citenamefont {Kwan}, \citenamefont {Henderson},
  \citenamefont {Gostic}, \citenamefont {Bredeweg}, \citenamefont {Couture},
  \citenamefont {Haight}, \citenamefont {Hayes-Sterbenz}, \citenamefont
  {Jandel}, \citenamefont {Lee}, \citenamefont {O'Donnell},\ and\ \citenamefont
  {Ullmann}}]{Chyzh2013}%
  \BibitemOpen
  \bibfield  {author} {\bibinfo {author} {\bibfnamefont {A.}~\bibnamefont
  {Chyzh}}, \bibinfo {author} {\bibfnamefont {C.~Y.}\ \bibnamefont {Wu}},
  \bibinfo {author} {\bibfnamefont {E.}~\bibnamefont {Kwan}}, \bibinfo {author}
  {\bibfnamefont {R.~A.}\ \bibnamefont {Henderson}}, \bibinfo {author}
  {\bibfnamefont {J.~M.}\ \bibnamefont {Gostic}}, \bibinfo {author}
  {\bibfnamefont {T.~A.}\ \bibnamefont {Bredeweg}}, \bibinfo {author}
  {\bibfnamefont {A.}~\bibnamefont {Couture}}, \bibinfo {author} {\bibfnamefont
  {R.~C.}\ \bibnamefont {Haight}}, \bibinfo {author} {\bibfnamefont {A.~C.}\
  \bibnamefont {Hayes-Sterbenz}}, \bibinfo {author} {\bibfnamefont
  {M.}~\bibnamefont {Jandel}}, \bibinfo {author} {\bibfnamefont {H.~Y.}\
  \bibnamefont {Lee}}, \bibinfo {author} {\bibfnamefont {J.~M.}\ \bibnamefont
  {O'Donnell}}, \ and\ \bibinfo {author} {\bibfnamefont {J.~L.}\ \bibnamefont
  {Ullmann}},\ }\href {\doibase 10.1103/PhysRevC.87.034620} {\bibfield
  {journal} {\bibinfo  {journal} {Phys. Rev. C}\ }\textbf {\bibinfo {volume}
  {87}},\ \bibinfo {pages} {034620} (\bibinfo {year} {2013})}\BibitemShut
  {NoStop}%
\bibitem [{\citenamefont {Gatera}\ \emph {et~al.}(2017)\citenamefont {Gatera},
  \citenamefont {Belgya}, \citenamefont {Geerts}, \citenamefont {G\"o\"ok},
  \citenamefont {Hambsch}, \citenamefont {Lebois}, \citenamefont {Mar\'oti},
  \citenamefont {Moens}, \citenamefont {Oberstedt}, \citenamefont {Oberstedt},
  \citenamefont {Postelt}, \citenamefont {Qi}, \citenamefont {Szentmikl\'osi},
  \citenamefont {Sibbens}, \citenamefont {Vanleeuw}, \citenamefont {Vidali},\
  and\ \citenamefont {Zeiser}}]{Gatera2017}%
  \BibitemOpen
  \bibfield  {author} {\bibinfo {author} {\bibfnamefont {A.}~\bibnamefont
  {Gatera}}, \bibinfo {author} {\bibfnamefont {T.}~\bibnamefont {Belgya}},
  \bibinfo {author} {\bibfnamefont {W.}~\bibnamefont {Geerts}}, \bibinfo
  {author} {\bibfnamefont {A.}~\bibnamefont {G\"o\"ok}}, \bibinfo {author}
  {\bibfnamefont {F.-J.}\ \bibnamefont {Hambsch}}, \bibinfo {author}
  {\bibfnamefont {M.}~\bibnamefont {Lebois}}, \bibinfo {author} {\bibfnamefont
  {B.}~\bibnamefont {Mar\'oti}}, \bibinfo {author} {\bibfnamefont
  {A.}~\bibnamefont {Moens}}, \bibinfo {author} {\bibfnamefont
  {A.}~\bibnamefont {Oberstedt}}, \bibinfo {author} {\bibfnamefont
  {S.}~\bibnamefont {Oberstedt}}, \bibinfo {author} {\bibfnamefont
  {F.}~\bibnamefont {Postelt}}, \bibinfo {author} {\bibfnamefont
  {L.}~\bibnamefont {Qi}}, \bibinfo {author} {\bibfnamefont {L.}~\bibnamefont
  {Szentmikl\'osi}}, \bibinfo {author} {\bibfnamefont {G.}~\bibnamefont
  {Sibbens}}, \bibinfo {author} {\bibfnamefont {D.}~\bibnamefont {Vanleeuw}},
  \bibinfo {author} {\bibfnamefont {M.}~\bibnamefont {Vidali}}, \ and\ \bibinfo
  {author} {\bibfnamefont {F.}~\bibnamefont {Zeiser}},\ }\href {\doibase
  10.1103/PhysRevC.95.064609} {\bibfield  {journal} {\bibinfo  {journal} {Phys.
  Rev. C}\ }\textbf {\bibinfo {volume} {95}},\ \bibinfo {pages} {064609}
  (\bibinfo {year} {2017})}\BibitemShut {NoStop}%
\bibitem [{\citenamefont {Pioro}(2016)}]{Pioro2016}%
  \BibitemOpen
  \bibfield  {author} {\bibinfo {author} {\bibfnamefont {I.}~\bibnamefont
  {Pioro}},\ }in\ \href {\doibase
  https://doi.org/10.1016/B978-0-08-100149-3.00002-1} {\emph {\bibinfo
  {booktitle} {Handbook of Generation \{IV\} Nuclear Reactors}}},\ \bibinfo
  {series and number} {Woodhead Publishing Series in Energy},\ \bibinfo
  {editor} {edited by\ \bibinfo {editor} {\bibfnamefont {I.~L.}\ \bibnamefont
  {Pioro}}}\ (\bibinfo  {publisher} {Woodhead Publishing},\ \bibinfo {year}
  {2016})\ pp.\ \bibinfo {pages} {37 -- 54}\BibitemShut {NoStop}%
\bibitem [{NEA()}]{NEAHPPu3}%
  \BibitemOpen
  \href {http://www.oecd-nea.org/dbdata/hprl/hprlview.pl?ID=421} {\enquote
  {\bibinfo {title} {{Nuclear Data High Priority Request List of the NEA} (req.
  id: H.3) http://www.oecd-nea.org/dbdata/hprl/hprlview.pl?id=421},}\
  }\BibitemShut {NoStop}%
\bibitem [{\citenamefont {Pelle}\ and\ \citenamefont
  {Maienschein}(1971)}]{Pelle1971}%
  \BibitemOpen
  \bibfield  {author} {\bibinfo {author} {\bibfnamefont {R.}~\bibnamefont
  {Pelle}}\ and\ \bibinfo {author} {\bibfnamefont {F.}~\bibnamefont
  {Maienschein}},\ }\href@noop {} {\bibfield  {journal} {\bibinfo  {journal}
  {Phys. Rev. C}\ ,\ \bibinfo {pages} {373}} (\bibinfo {year}
  {1971})}\BibitemShut {NoStop}%
\bibitem [{\citenamefont {Verbinski}\ \emph {et~al.}(1973)\citenamefont
  {Verbinski}, \citenamefont {Weber},\ and\ \citenamefont
  {Sund}}]{verbinski1973prompt}%
  \BibitemOpen
  \bibfield  {author} {\bibinfo {author} {\bibfnamefont {V.}~\bibnamefont
  {Verbinski}}, \bibinfo {author} {\bibfnamefont {H.}~\bibnamefont {Weber}}, \
  and\ \bibinfo {author} {\bibfnamefont {R.}~\bibnamefont {Sund}},\ }\href@noop
  {} {\bibfield  {journal} {\bibinfo  {journal} {Physical Review C}\ }\textbf
  {\bibinfo {volume} {7}},\ \bibinfo {pages} {1173} (\bibinfo {year}
  {1973})}\BibitemShut {NoStop}%
\bibitem [{\citenamefont {Pleasonton}(1973)}]{pleasonton1973prompt}%
  \BibitemOpen
  \bibfield  {author} {\bibinfo {author} {\bibfnamefont {F.}~\bibnamefont
  {Pleasonton}},\ }\href@noop {} {\bibfield  {journal} {\bibinfo  {journal}
  {Nuclear Physics A}\ }\textbf {\bibinfo {volume} {213}},\ \bibinfo {pages}
  {413} (\bibinfo {year} {1973})}\BibitemShut {NoStop}%
\bibitem [{\citenamefont {{M. Salvatores et al.}}(2008)}]{Salvatores2008}%
  \BibitemOpen
  \bibfield  {author} {\bibinfo {author} {\bibnamefont {{M. Salvatores et
  al.}}},\ }\href {https://www.oecd-nea.org/science/wpec/volume26/volume26.pdf}
  {\emph {\bibinfo {title} {Uncertainty and target accuracy assessment for
  innovative systems using recent covariance data evaluations}}},\ \bibinfo
  {type} {Tech. Rep.}\ \bibinfo {number} {Volume 26}\ (\bibinfo  {institution}
  {OECD/NEA WPEC},\ \bibinfo {year} {2008})\BibitemShut {NoStop}%
\bibitem [{\citenamefont {Billnert}\ \emph {et~al.}(2013)\citenamefont
  {Billnert}, \citenamefont {Hambsch}, \citenamefont {Oberstedt},\ and\
  \citenamefont {Oberstedt}}]{billnert2013new}%
  \BibitemOpen
  \bibfield  {author} {\bibinfo {author} {\bibfnamefont {R.}~\bibnamefont
  {Billnert}}, \bibinfo {author} {\bibfnamefont {F.-J.}\ \bibnamefont
  {Hambsch}}, \bibinfo {author} {\bibfnamefont {A.}~\bibnamefont {Oberstedt}},
  \ and\ \bibinfo {author} {\bibfnamefont {S.}~\bibnamefont {Oberstedt}},\
  }\href@noop {} {\bibfield  {journal} {\bibinfo  {journal} {Physical Review
  C}\ }\textbf {\bibinfo {volume} {87}},\ \bibinfo {pages} {024601} (\bibinfo
  {year} {2013})}\BibitemShut {NoStop}%
\bibitem [{\citenamefont {Cramer}\ and\ \citenamefont
  {Britt}(1970)}]{cramer1970neutron}%
  \BibitemOpen
  \bibfield  {author} {\bibinfo {author} {\bibfnamefont {J.}~\bibnamefont
  {Cramer}}\ and\ \bibinfo {author} {\bibfnamefont {H.}~\bibnamefont {Britt}},\
  }\href@noop {} {\bibfield  {journal} {\bibinfo  {journal} {Nuclear Science
  and Engineering}\ }\textbf {\bibinfo {volume} {41}},\ \bibinfo {pages} {177}
  (\bibinfo {year} {1970})}\BibitemShut {NoStop}%
\bibitem [{\citenamefont {Guttormsen}\ \emph {et~al.}(1990)\citenamefont
  {Guttormsen}, \citenamefont {Atac}, \citenamefont {L{\o}vh{\o}iden},
  \citenamefont {Messelt}, \citenamefont {Rams{\o}y}, \citenamefont {Rekstad},
  \citenamefont {Thorsteinsen}, \citenamefont {Tveter},\ and\ \citenamefont
  {Zelazny}}]{guttormsen1990statistical}%
  \BibitemOpen
  \bibfield  {author} {\bibinfo {author} {\bibfnamefont {M.}~\bibnamefont
  {Guttormsen}}, \bibinfo {author} {\bibfnamefont {A.}~\bibnamefont {Atac}},
  \bibinfo {author} {\bibfnamefont {G.}~\bibnamefont {L{\o}vh{\o}iden}},
  \bibinfo {author} {\bibfnamefont {S.}~\bibnamefont {Messelt}}, \bibinfo
  {author} {\bibfnamefont {T.}~\bibnamefont {Rams{\o}y}}, \bibinfo {author}
  {\bibfnamefont {J.}~\bibnamefont {Rekstad}}, \bibinfo {author} {\bibfnamefont
  {T.}~\bibnamefont {Thorsteinsen}}, \bibinfo {author} {\bibfnamefont
  {T.}~\bibnamefont {Tveter}}, \ and\ \bibinfo {author} {\bibfnamefont
  {Z.}~\bibnamefont {Zelazny}},\ }\href@noop {} {\bibfield  {journal} {\bibinfo
   {journal} {Physica Scripta}\ }\textbf {\bibinfo {volume} {1990}},\ \bibinfo
  {pages} {54} (\bibinfo {year} {1990})}\BibitemShut {NoStop}%
\bibitem [{\citenamefont {Guttormsen}\ \emph {et~al.}(2011)\citenamefont
  {Guttormsen}, \citenamefont {B{\"u}rger}, \citenamefont {Hansen},\ and\
  \citenamefont {Lietaer}}]{guttormsen2011siri}%
  \BibitemOpen
  \bibfield  {author} {\bibinfo {author} {\bibfnamefont {M.}~\bibnamefont
  {Guttormsen}}, \bibinfo {author} {\bibfnamefont {A.}~\bibnamefont
  {B{\"u}rger}}, \bibinfo {author} {\bibfnamefont {T.}~\bibnamefont {Hansen}},
  \ and\ \bibinfo {author} {\bibfnamefont {N.}~\bibnamefont {Lietaer}},\
  }\href@noop {} {\bibfield  {journal} {\bibinfo  {journal} {Nuclear
  Instruments and Methods in Physics Research Section A: Accelerators,
  Spectrometers, Detectors and Associated Equipment}\ }\textbf {\bibinfo
  {volume} {648}},\ \bibinfo {pages} {168} (\bibinfo {year}
  {2011})}\BibitemShut {NoStop}%
\bibitem [{\citenamefont {Tornyi}\ \emph {et~al.}(2014)\citenamefont {Tornyi},
  \citenamefont {G{\"o}rgen}, \citenamefont {Guttormsen}, \citenamefont
  {Larsen}, \citenamefont {Siem}, \citenamefont {Krasznahorkay},\ and\
  \citenamefont {Csige}}]{tornyi2014new}%
  \BibitemOpen
  \bibfield  {author} {\bibinfo {author} {\bibfnamefont {T.~G.}\ \bibnamefont
  {Tornyi}}, \bibinfo {author} {\bibfnamefont {A.}~\bibnamefont {G{\"o}rgen}},
  \bibinfo {author} {\bibfnamefont {M.}~\bibnamefont {Guttormsen}}, \bibinfo
  {author} {\bibfnamefont {A.-C.}\ \bibnamefont {Larsen}}, \bibinfo {author}
  {\bibfnamefont {S.}~\bibnamefont {Siem}}, \bibinfo {author} {\bibfnamefont
  {A.}~\bibnamefont {Krasznahorkay}}, \ and\ \bibinfo {author} {\bibfnamefont
  {L.}~\bibnamefont {Csige}},\ }\href@noop {} {\bibfield  {journal} {\bibinfo
  {journal} {Nuclear Instruments and Methods in Physics Research Section A:
  Accelerators, Spectrometers, Detectors and Associated Equipment}\ }\textbf
  {\bibinfo {volume} {738}},\ \bibinfo {pages} {6} (\bibinfo {year}
  {2014})}\BibitemShut {NoStop}%
\bibitem [{\citenamefont {Henderson}\ \emph {et~al.}(2011)\citenamefont
  {Henderson}, \citenamefont {Gostic}, \citenamefont {Burke}, \citenamefont
  {Fisher},\ and\ \citenamefont {Wu}}]{Henderson2011}%
  \BibitemOpen
  \bibfield  {author} {\bibinfo {author} {\bibfnamefont {R.}~\bibnamefont
  {Henderson}}, \bibinfo {author} {\bibfnamefont {J.}~\bibnamefont {Gostic}},
  \bibinfo {author} {\bibfnamefont {J.}~\bibnamefont {Burke}}, \bibinfo
  {author} {\bibfnamefont {S.}~\bibnamefont {Fisher}}, \ and\ \bibinfo {author}
  {\bibfnamefont {C.}~\bibnamefont {Wu}},\ }\href {\doibase
  10.1016/j.nima.2011.06.023} {\bibfield  {journal} {\bibinfo  {journal}
  {Nuclear Instruments and Methods in Physics Research Section A: Accelerators,
  Spectrometers, Detectors and Associated Equipment}\ }\textbf {\bibinfo
  {volume} {655}},\ \bibinfo {pages} {66–71} (\bibinfo {year}
  {2011})}\BibitemShut {NoStop}%
\bibitem [{\citenamefont {Capote}\ \emph {et~al.}(2009)\citenamefont {Capote},
  \citenamefont {Herman}, \citenamefont {Oblo{\v{z}}insk{\`y}}, \citenamefont
  {Young}, \citenamefont {Goriely}, \citenamefont {Belgya}, \citenamefont
  {Ignatyuk}, \citenamefont {Koning}, \citenamefont {Hilaire}, \citenamefont
  {Plujko} \emph {et~al.}}]{capote2009ripl}%
  \BibitemOpen
  \bibfield  {author} {\bibinfo {author} {\bibfnamefont {R.}~\bibnamefont
  {Capote}}, \bibinfo {author} {\bibfnamefont {M.}~\bibnamefont {Herman}},
  \bibinfo {author} {\bibfnamefont {P.}~\bibnamefont {Oblo{\v{z}}insk{\`y}}},
  \bibinfo {author} {\bibfnamefont {P.}~\bibnamefont {Young}}, \bibinfo
  {author} {\bibfnamefont {S.}~\bibnamefont {Goriely}}, \bibinfo {author}
  {\bibfnamefont {T.}~\bibnamefont {Belgya}}, \bibinfo {author} {\bibfnamefont
  {A.}~\bibnamefont {Ignatyuk}}, \bibinfo {author} {\bibfnamefont {A.~J.}\
  \bibnamefont {Koning}}, \bibinfo {author} {\bibfnamefont {S.}~\bibnamefont
  {Hilaire}}, \bibinfo {author} {\bibfnamefont {V.~A.}\ \bibnamefont {Plujko}},
   \emph {et~al.},\ }\href@noop {} {\bibfield  {journal} {\bibinfo  {journal}
  {Nuclear Data Sheets}\ }\textbf {\bibinfo {volume} {110}},\ \bibinfo {pages}
  {3107} (\bibinfo {year} {2009})}\BibitemShut {NoStop}%
\bibitem [{\citenamefont {Ducasse}\ \emph {et~al.}(2015)\citenamefont
  {Ducasse}, \citenamefont {Jurado}, \citenamefont {A{\"\i}che}, \citenamefont
  {Marini}, \citenamefont {Mathieu}, \citenamefont {G{\"o}rgen}, \citenamefont
  {Guttormsen}, \citenamefont {Larsen}, \citenamefont {Tornyi}, \citenamefont
  {Wilson} \emph {et~al.}}]{ducasse2015study}%
  \BibitemOpen
  \bibfield  {author} {\bibinfo {author} {\bibfnamefont {Q.}~\bibnamefont
  {Ducasse}}, \bibinfo {author} {\bibfnamefont {B.}~\bibnamefont {Jurado}},
  \bibinfo {author} {\bibfnamefont {M.}~\bibnamefont {A{\"\i}che}}, \bibinfo
  {author} {\bibfnamefont {P.}~\bibnamefont {Marini}}, \bibinfo {author}
  {\bibfnamefont {L.}~\bibnamefont {Mathieu}}, \bibinfo {author} {\bibfnamefont
  {A.}~\bibnamefont {G{\"o}rgen}}, \bibinfo {author} {\bibfnamefont
  {M.}~\bibnamefont {Guttormsen}}, \bibinfo {author} {\bibfnamefont
  {A.}~\bibnamefont {Larsen}}, \bibinfo {author} {\bibfnamefont
  {T.}~\bibnamefont {Tornyi}}, \bibinfo {author} {\bibfnamefont
  {J.}~\bibnamefont {Wilson}},  \emph {et~al.},\ }\href@noop {} {\bibfield
  {journal} {\bibinfo  {journal} {arXiv preprint arXiv:1512.06334}\ } (\bibinfo
  {year} {2015})}\BibitemShut {NoStop}%
\bibitem [{\citenamefont {Guttormsen}\ \emph {et~al.}(1996)\citenamefont
  {Guttormsen}, \citenamefont {Tveter}, \citenamefont {Bergholt}, \citenamefont
  {Ingebretsen},\ and\ \citenamefont {Rekstad}}]{guttormsen1996unfolding}%
  \BibitemOpen
  \bibfield  {author} {\bibinfo {author} {\bibfnamefont {M.}~\bibnamefont
  {Guttormsen}}, \bibinfo {author} {\bibfnamefont {T.}~\bibnamefont {Tveter}},
  \bibinfo {author} {\bibfnamefont {L.}~\bibnamefont {Bergholt}}, \bibinfo
  {author} {\bibfnamefont {F.}~\bibnamefont {Ingebretsen}}, \ and\ \bibinfo
  {author} {\bibfnamefont {J.}~\bibnamefont {Rekstad}},\ }\href@noop {}
  {\bibfield  {journal} {\bibinfo  {journal} {Nuclear Instruments and Methods
  in Physics Research Section A: Accelerators, Spectrometers, Detectors and
  Associated Equipment}\ }\textbf {\bibinfo {volume} {374}},\ \bibinfo {pages}
  {371} (\bibinfo {year} {1996})}\BibitemShut {NoStop}%
\bibitem [{\citenamefont {Crespo~Campo}\ \emph {et~al.}(2016)\citenamefont
  {Crespo~Campo}, \citenamefont {Bello~Garrote}, \citenamefont {Eriksen},
  \citenamefont {G\"orgen}, \citenamefont {Guttormsen}, \citenamefont
  {Hadynska-Klek}, \citenamefont {Klintefjord}, \citenamefont {Larsen},
  \citenamefont {Renstr\o{}m}, \citenamefont {Sahin}, \citenamefont {Siem},
  \citenamefont {Springer}, \citenamefont {Tornyi},\ and\ \citenamefont
  {Tveten}}]{PhysRevC.94.044321}%
  \BibitemOpen
  \bibfield  {author} {\bibinfo {author} {\bibfnamefont {L.}~\bibnamefont
  {Crespo~Campo}}, \bibinfo {author} {\bibfnamefont {F.~L.}\ \bibnamefont
  {Bello~Garrote}}, \bibinfo {author} {\bibfnamefont {T.~K.}\ \bibnamefont
  {Eriksen}}, \bibinfo {author} {\bibfnamefont {A.}~\bibnamefont {G\"orgen}},
  \bibinfo {author} {\bibfnamefont {M.}~\bibnamefont {Guttormsen}}, \bibinfo
  {author} {\bibfnamefont {K.}~\bibnamefont {Hadynska-Klek}}, \bibinfo {author}
  {\bibfnamefont {M.}~\bibnamefont {Klintefjord}}, \bibinfo {author}
  {\bibfnamefont {A.~C.}\ \bibnamefont {Larsen}}, \bibinfo {author}
  {\bibfnamefont {T.}~\bibnamefont {Renstr\o{}m}}, \bibinfo {author}
  {\bibfnamefont {E.}~\bibnamefont {Sahin}}, \bibinfo {author} {\bibfnamefont
  {S.}~\bibnamefont {Siem}}, \bibinfo {author} {\bibfnamefont {A.}~\bibnamefont
  {Springer}}, \bibinfo {author} {\bibfnamefont {T.~G.}\ \bibnamefont
  {Tornyi}}, \ and\ \bibinfo {author} {\bibfnamefont {G.~M.}\ \bibnamefont
  {Tveten}},\ }\href {\doibase 10.1103/PhysRevC.94.044321} {\bibfield
  {journal} {\bibinfo  {journal} {Phys. Rev. C}\ }\textbf {\bibinfo {volume}
  {94}},\ \bibinfo {pages} {044321} (\bibinfo {year} {2016})}\BibitemShut
  {NoStop}%
\bibitem [{\citenamefont {H{\"a}usser}\ \emph {et~al.}(1983)\citenamefont
  {H{\"a}usser}, \citenamefont {Lone}, \citenamefont {Alexander}, \citenamefont
  {Kushneriuk},\ and\ \citenamefont {Gascon}}]{hausser1983prompt}%
  \BibitemOpen
  \bibfield  {author} {\bibinfo {author} {\bibfnamefont {O.}~\bibnamefont
  {H{\"a}usser}}, \bibinfo {author} {\bibfnamefont {M.}~\bibnamefont {Lone}},
  \bibinfo {author} {\bibfnamefont {T.}~\bibnamefont {Alexander}}, \bibinfo
  {author} {\bibfnamefont {S.}~\bibnamefont {Kushneriuk}}, \ and\ \bibinfo
  {author} {\bibfnamefont {J.}~\bibnamefont {Gascon}},\ }\href@noop {}
  {\bibfield  {journal} {\bibinfo  {journal} {Nuclear Instruments and Methods
  in Physics Research}\ }\textbf {\bibinfo {volume} {213}},\ \bibinfo {pages}
  {301} (\bibinfo {year} {1983})}\BibitemShut {NoStop}%
\bibitem [{\citenamefont {Ullmann}\ \emph {et~al.}(2013)\citenamefont
  {Ullmann}, \citenamefont {Bond}, \citenamefont {Bredeweg}, \citenamefont
  {Couture}, \citenamefont {Haight}, \citenamefont {Jandel}, \citenamefont
  {Kawano}, \citenamefont {Lee}, \citenamefont {O’Donnell}, \citenamefont
  {Hayes} \emph {et~al.}}]{ullmann2013prompt}%
  \BibitemOpen
  \bibfield  {author} {\bibinfo {author} {\bibfnamefont {J.}~\bibnamefont
  {Ullmann}}, \bibinfo {author} {\bibfnamefont {E.}~\bibnamefont {Bond}},
  \bibinfo {author} {\bibfnamefont {T.}~\bibnamefont {Bredeweg}}, \bibinfo
  {author} {\bibfnamefont {A.}~\bibnamefont {Couture}}, \bibinfo {author}
  {\bibfnamefont {R.}~\bibnamefont {Haight}}, \bibinfo {author} {\bibfnamefont
  {M.}~\bibnamefont {Jandel}}, \bibinfo {author} {\bibfnamefont
  {T.}~\bibnamefont {Kawano}}, \bibinfo {author} {\bibfnamefont
  {H.}~\bibnamefont {Lee}}, \bibinfo {author} {\bibfnamefont {J.}~\bibnamefont
  {O’Donnell}}, \bibinfo {author} {\bibfnamefont {A.}~\bibnamefont {Hayes}},
  \emph {et~al.},\ }\href@noop {} {\bibfield  {journal} {\bibinfo  {journal}
  {Physical Review C}\ }\textbf {\bibinfo {volume} {87}},\ \bibinfo {pages}
  {044607} (\bibinfo {year} {2013})}\BibitemShut {NoStop}%
\bibitem [{END(2011{\natexlab{a}})}]{ENDF233U}%
  \BibitemOpen
  \href@noop {} {\bibfield  {journal} {\bibinfo  {journal} {ENDF/B-VII.1
  Evaluated Nuclear Data Files, ZA=92233, NSUB=10(N), MT=456}\ ,\ \bibinfo
  {pages} {www.nndc.bnl.gov/exfor/endf02.jsp}} (\bibinfo {year}
  {2011}{\natexlab{a}})}\BibitemShut {NoStop}%
\bibitem [{END(2011{\natexlab{b}})}]{ENDF239Pu}%
  \BibitemOpen
  \href@noop {} {\bibfield  {journal} {\bibinfo  {journal} {ENDF/B-VII.1
  Evaluated Nuclear Data Files, ZA=94239, NSUB=10(N), MT=456}\ ,\ \bibinfo
  {pages} {www.nndc.bnl.gov/exfor/endf02.jsp}} (\bibinfo {year}
  {2011}{\natexlab{b}})}\BibitemShut {NoStop}%
\bibitem [{\citenamefont {Madland}(2006)}]{madland2006total}%
  \BibitemOpen
  \bibfield  {author} {\bibinfo {author} {\bibfnamefont {D.}~\bibnamefont
  {Madland}},\ }\href@noop {} {\bibfield  {journal} {\bibinfo  {journal}
  {Nuclear Physics A}\ }\textbf {\bibinfo {volume} {772}},\ \bibinfo {pages}
  {113} (\bibinfo {year} {2006})}\BibitemShut {NoStop}%
\bibitem [{\citenamefont {Chadwick}\ \emph {et~al.}(2011)\citenamefont
  {Chadwick}, \citenamefont {Herman}, \citenamefont {Oblo{\v{z}}insk{\`y}},
  \citenamefont {Dunn}, \citenamefont {Danon}, \citenamefont {Kahler},
  \citenamefont {Smith}, \citenamefont {Pritychenko}, \citenamefont {Arbanas},
  \citenamefont {Arcilla} \emph {et~al.}}]{chadwick2011endf}%
  \BibitemOpen
  \bibfield  {author} {\bibinfo {author} {\bibfnamefont {M.}~\bibnamefont
  {Chadwick}}, \bibinfo {author} {\bibfnamefont {M.}~\bibnamefont {Herman}},
  \bibinfo {author} {\bibfnamefont {P.}~\bibnamefont {Oblo{\v{z}}insk{\`y}}},
  \bibinfo {author} {\bibfnamefont {M.~E.}\ \bibnamefont {Dunn}}, \bibinfo
  {author} {\bibfnamefont {Y.}~\bibnamefont {Danon}}, \bibinfo {author}
  {\bibfnamefont {A.}~\bibnamefont {Kahler}}, \bibinfo {author} {\bibfnamefont
  {D.~L.}\ \bibnamefont {Smith}}, \bibinfo {author} {\bibfnamefont
  {B.}~\bibnamefont {Pritychenko}}, \bibinfo {author} {\bibfnamefont
  {G.}~\bibnamefont {Arbanas}}, \bibinfo {author} {\bibfnamefont
  {R.}~\bibnamefont {Arcilla}},  \emph {et~al.},\ }\href@noop {} {\bibfield
  {journal} {\bibinfo  {journal} {Nuclear Data Sheets}\ }\textbf {\bibinfo
  {volume} {112}},\ \bibinfo {pages} {2887} (\bibinfo {year}
  {2011})}\BibitemShut {NoStop}%
\bibitem [{\citenamefont {Chyzh}\ \emph {et~al.}(2012)\citenamefont {Chyzh},
  \citenamefont {Wu}, \citenamefont {Kwan}, \citenamefont {Henderson},
  \citenamefont {Gostic}, \citenamefont {Bredeweg}, \citenamefont {Haight},
  \citenamefont {Hayes-Sterbenz}, \citenamefont {Jandel}, \citenamefont
  {O’Donnell} \emph {et~al.}}]{chyzh2012evidence}%
  \BibitemOpen
  \bibfield  {author} {\bibinfo {author} {\bibfnamefont {A.}~\bibnamefont
  {Chyzh}}, \bibinfo {author} {\bibfnamefont {C.}~\bibnamefont {Wu}}, \bibinfo
  {author} {\bibfnamefont {E.}~\bibnamefont {Kwan}}, \bibinfo {author}
  {\bibfnamefont {R.}~\bibnamefont {Henderson}}, \bibinfo {author}
  {\bibfnamefont {J.}~\bibnamefont {Gostic}}, \bibinfo {author} {\bibfnamefont
  {T.}~\bibnamefont {Bredeweg}}, \bibinfo {author} {\bibfnamefont
  {R.}~\bibnamefont {Haight}}, \bibinfo {author} {\bibfnamefont
  {A.}~\bibnamefont {Hayes-Sterbenz}}, \bibinfo {author} {\bibfnamefont
  {M.}~\bibnamefont {Jandel}}, \bibinfo {author} {\bibfnamefont
  {J.}~\bibnamefont {O’Donnell}},  \emph {et~al.},\ }\href@noop {} {\bibfield
   {journal} {\bibinfo  {journal} {Physical Review C}\ }\textbf {\bibinfo
  {volume} {85}},\ \bibinfo {pages} {021601} (\bibinfo {year}
  {2012})}\BibitemShut {NoStop}%
\bibitem [{\citenamefont {Lebois}\ \emph {et~al.}(2015)\citenamefont {Lebois},
  \citenamefont {Wilson}, \citenamefont {Halipr{\'e}}, \citenamefont
  {Oberstedt}, \citenamefont {Oberstedt}, \citenamefont {Marini}, \citenamefont
  {Schmitt}, \citenamefont {Rose}, \citenamefont {Siem}, \citenamefont {Fallot}
  \emph {et~al.}}]{lebois2015comparative}%
  \BibitemOpen
  \bibfield  {author} {\bibinfo {author} {\bibfnamefont {M.}~\bibnamefont
  {Lebois}}, \bibinfo {author} {\bibfnamefont {J.}~\bibnamefont {Wilson}},
  \bibinfo {author} {\bibfnamefont {P.}~\bibnamefont {Halipr{\'e}}}, \bibinfo
  {author} {\bibfnamefont {A.}~\bibnamefont {Oberstedt}}, \bibinfo {author}
  {\bibfnamefont {S.}~\bibnamefont {Oberstedt}}, \bibinfo {author}
  {\bibfnamefont {P.}~\bibnamefont {Marini}}, \bibinfo {author} {\bibfnamefont
  {C.}~\bibnamefont {Schmitt}}, \bibinfo {author} {\bibfnamefont
  {S.}~\bibnamefont {Rose}}, \bibinfo {author} {\bibfnamefont {S.}~\bibnamefont
  {Siem}}, \bibinfo {author} {\bibfnamefont {M.}~\bibnamefont {Fallot}},  \emph
  {et~al.},\ }\href@noop {} {\bibfield  {journal} {\bibinfo  {journal}
  {Physical Review C}\ }\textbf {\bibinfo {volume} {92}},\ \bibinfo {pages}
  {034618} (\bibinfo {year} {2015})}\BibitemShut {NoStop}%
\bibitem [{\citenamefont {Schmidt}\ \emph {et~al.}(2016)\citenamefont
  {Schmidt}, \citenamefont {Jurado}, \citenamefont {Amouroux},\ and\
  \citenamefont {Schmitt}}]{schmidt2016general}%
  \BibitemOpen
  \bibfield  {author} {\bibinfo {author} {\bibfnamefont {K.-H.}\ \bibnamefont
  {Schmidt}}, \bibinfo {author} {\bibfnamefont {B.}~\bibnamefont {Jurado}},
  \bibinfo {author} {\bibfnamefont {C.}~\bibnamefont {Amouroux}}, \ and\
  \bibinfo {author} {\bibfnamefont {C.}~\bibnamefont {Schmitt}},\ }\href@noop
  {} {\bibfield  {journal} {\bibinfo  {journal} {Nuclear Data Sheets}\ }\textbf
  {\bibinfo {volume} {131}},\ \bibinfo {pages} {107} (\bibinfo {year}
  {2016})}\BibitemShut {NoStop}%
\bibitem [{\citenamefont {Ericson}(1960)}]{Ericson1960}%
  \BibitemOpen
  \bibfield  {author} {\bibinfo {author} {\bibfnamefont {T.}~\bibnamefont
  {Ericson}},\ }\href {\doibase 10.1080/00018736000101239} {\bibfield
  {journal} {\bibinfo  {journal} {Advances in Physics}\ }\textbf {\bibinfo
  {volume} {9}},\ \bibinfo {pages} {425–511} (\bibinfo {year}
  {1960})}\BibitemShut {NoStop}%
\bibitem [{\citenamefont {Guttormsen}\ \emph {et~al.}(2013)\citenamefont
  {Guttormsen}, \citenamefont {Jurado}, \citenamefont {Wilson}, \citenamefont
  {Aiche}, \citenamefont {Bernstein}, \citenamefont {Ducasse}, \citenamefont
  {Giacoppo}, \citenamefont {G{\"o}rgen}, \citenamefont {Gunsing},
  \citenamefont {Hagen} \emph {et~al.}}]{guttormsen2013constant}%
  \BibitemOpen
  \bibfield  {author} {\bibinfo {author} {\bibfnamefont {M.}~\bibnamefont
  {Guttormsen}}, \bibinfo {author} {\bibfnamefont {B.}~\bibnamefont {Jurado}},
  \bibinfo {author} {\bibfnamefont {J.}~\bibnamefont {Wilson}}, \bibinfo
  {author} {\bibfnamefont {M.}~\bibnamefont {Aiche}}, \bibinfo {author}
  {\bibfnamefont {L.}~\bibnamefont {Bernstein}}, \bibinfo {author}
  {\bibfnamefont {Q.}~\bibnamefont {Ducasse}}, \bibinfo {author} {\bibfnamefont
  {F.}~\bibnamefont {Giacoppo}}, \bibinfo {author} {\bibfnamefont
  {A.}~\bibnamefont {G{\"o}rgen}}, \bibinfo {author} {\bibfnamefont
  {F.}~\bibnamefont {Gunsing}}, \bibinfo {author} {\bibfnamefont
  {T.}~\bibnamefont {Hagen}},  \emph {et~al.},\ }\href@noop {} {\bibfield
  {journal} {\bibinfo  {journal} {Physical Review C}\ }\textbf {\bibinfo
  {volume} {88}},\ \bibinfo {pages} {024307} (\bibinfo {year}
  {2013})}\BibitemShut {NoStop}%
\bibitem [{\citenamefont {Guttormsen}\ \emph {et~al.}(2014)\citenamefont
  {Guttormsen}, \citenamefont {Bernstein}, \citenamefont {G{\"o}rgen},
  \citenamefont {Jurado}, \citenamefont {Siem}, \citenamefont {Aiche},
  \citenamefont {Ducasse}, \citenamefont {Giacoppo}, \citenamefont {Gunsing},
  \citenamefont {Hagen} \emph {et~al.}}]{guttormsen2014scissors}%
  \BibitemOpen
  \bibfield  {author} {\bibinfo {author} {\bibfnamefont {M.}~\bibnamefont
  {Guttormsen}}, \bibinfo {author} {\bibfnamefont {L.}~\bibnamefont
  {Bernstein}}, \bibinfo {author} {\bibfnamefont {A.}~\bibnamefont
  {G{\"o}rgen}}, \bibinfo {author} {\bibfnamefont {B.}~\bibnamefont {Jurado}},
  \bibinfo {author} {\bibfnamefont {S.}~\bibnamefont {Siem}}, \bibinfo {author}
  {\bibfnamefont {M.}~\bibnamefont {Aiche}}, \bibinfo {author} {\bibfnamefont
  {Q.}~\bibnamefont {Ducasse}}, \bibinfo {author} {\bibfnamefont
  {F.}~\bibnamefont {Giacoppo}}, \bibinfo {author} {\bibfnamefont
  {F.}~\bibnamefont {Gunsing}}, \bibinfo {author} {\bibfnamefont
  {T.}~\bibnamefont {Hagen}},  \emph {et~al.},\ }\href@noop {} {\bibfield
  {journal} {\bibinfo  {journal} {Physical Review C}\ }\textbf {\bibinfo
  {volume} {89}},\ \bibinfo {pages} {014302} (\bibinfo {year}
  {2014})}\BibitemShut {NoStop}%
\bibitem [{\citenamefont {Bucurescu}\ and\ \citenamefont {von
  Egidy}(2005)}]{bucurescu2005systematics}%
  \BibitemOpen
  \bibfield  {author} {\bibinfo {author} {\bibfnamefont {D.}~\bibnamefont
  {Bucurescu}}\ and\ \bibinfo {author} {\bibfnamefont {T.}~\bibnamefont {von
  Egidy}},\ }\href@noop {} {\bibfield  {journal} {\bibinfo  {journal} {Journal
  of Physics G: Nuclear and Particle Physics}\ }\textbf {\bibinfo {volume}
  {31}},\ \bibinfo {pages} {S1675} (\bibinfo {year} {2005})}\BibitemShut
  {NoStop}%
\bibitem [{\citenamefont {Browne}\ and\ \citenamefont
  {Tuli}(2007)}]{browne2007nuclear}%
  \BibitemOpen
  \bibfield  {author} {\bibinfo {author} {\bibfnamefont {E.}~\bibnamefont
  {Browne}}\ and\ \bibinfo {author} {\bibfnamefont {J.}~\bibnamefont {Tuli}},\
  }\href@noop {} {\bibfield  {journal} {\bibinfo  {journal} {Nuclear Data
  Sheets}\ }\textbf {\bibinfo {volume} {108}},\ \bibinfo {pages} {681}
  (\bibinfo {year} {2007})}\BibitemShut {NoStop}%
\bibitem [{\citenamefont {Hunyadi}\ \emph {et~al.}(2001)\citenamefont
  {Hunyadi}, \citenamefont {Gassmann}, \citenamefont {Krasznahorkay},
  \citenamefont {Habs}, \citenamefont {Thirolf}, \citenamefont {Csatl{\'o}s},
  \citenamefont {Eisermann}, \citenamefont {Faestermann}, \citenamefont {Graw},
  \citenamefont {Guly{\'a}s} \emph {et~al.}}]{hunyadi2001excited}%
  \BibitemOpen
  \bibfield  {author} {\bibinfo {author} {\bibfnamefont {M.}~\bibnamefont
  {Hunyadi}}, \bibinfo {author} {\bibfnamefont {D.}~\bibnamefont {Gassmann}},
  \bibinfo {author} {\bibfnamefont {A.}~\bibnamefont {Krasznahorkay}}, \bibinfo
  {author} {\bibfnamefont {D.}~\bibnamefont {Habs}}, \bibinfo {author}
  {\bibfnamefont {P.}~\bibnamefont {Thirolf}}, \bibinfo {author} {\bibfnamefont
  {M.}~\bibnamefont {Csatl{\'o}s}}, \bibinfo {author} {\bibfnamefont
  {Y.}~\bibnamefont {Eisermann}}, \bibinfo {author} {\bibfnamefont
  {T.}~\bibnamefont {Faestermann}}, \bibinfo {author} {\bibfnamefont
  {G.}~\bibnamefont {Graw}}, \bibinfo {author} {\bibfnamefont {J.}~\bibnamefont
  {Guly{\'a}s}},  \emph {et~al.},\ }\href@noop {} {\bibfield  {journal}
  {\bibinfo  {journal} {Physics Letters B}\ }\textbf {\bibinfo {volume}
  {505}},\ \bibinfo {pages} {27} (\bibinfo {year} {2001})}\BibitemShut
  {NoStop}%
\bibitem [{\citenamefont {Gl{\"a}ssel}\ \emph {et~al.}(1976)\citenamefont
  {Gl{\"a}ssel}, \citenamefont {R{\"o}sler},\ and\ \citenamefont
  {Specht}}]{glassel1976intermediate}%
  \BibitemOpen
  \bibfield  {author} {\bibinfo {author} {\bibfnamefont {P.}~\bibnamefont
  {Gl{\"a}ssel}}, \bibinfo {author} {\bibfnamefont {H.}~\bibnamefont
  {R{\"o}sler}}, \ and\ \bibinfo {author} {\bibfnamefont {H.}~\bibnamefont
  {Specht}},\ }\href@noop {} {\bibfield  {journal} {\bibinfo  {journal}
  {Nuclear Physics A}\ }\textbf {\bibinfo {volume} {256}},\ \bibinfo {pages}
  {220} (\bibinfo {year} {1976})}\BibitemShut {NoStop}%
\bibitem [{\citenamefont {Boutoux}\ \emph {et~al.}(2012)\citenamefont
  {Boutoux}, \citenamefont {Jurado}, \citenamefont {M{\'e}ot}, \citenamefont
  {Roig}, \citenamefont {Mathieu}, \citenamefont {A{\"\i}che}, \citenamefont
  {Barreau}, \citenamefont {Capellan}, \citenamefont {Companis}, \citenamefont
  {Czajkowski} \emph {et~al.}}]{boutoux2012study}%
  \BibitemOpen
  \bibfield  {author} {\bibinfo {author} {\bibfnamefont {G.}~\bibnamefont
  {Boutoux}}, \bibinfo {author} {\bibfnamefont {B.}~\bibnamefont {Jurado}},
  \bibinfo {author} {\bibfnamefont {V.}~\bibnamefont {M{\'e}ot}}, \bibinfo
  {author} {\bibfnamefont {O.}~\bibnamefont {Roig}}, \bibinfo {author}
  {\bibfnamefont {L.}~\bibnamefont {Mathieu}}, \bibinfo {author} {\bibfnamefont
  {M.}~\bibnamefont {A{\"\i}che}}, \bibinfo {author} {\bibfnamefont
  {G.}~\bibnamefont {Barreau}}, \bibinfo {author} {\bibfnamefont
  {N.}~\bibnamefont {Capellan}}, \bibinfo {author} {\bibfnamefont
  {I.}~\bibnamefont {Companis}}, \bibinfo {author} {\bibfnamefont
  {S.}~\bibnamefont {Czajkowski}},  \emph {et~al.},\ }\href@noop {} {\bibfield
  {journal} {\bibinfo  {journal} {Physics Letters B}\ }\textbf {\bibinfo
  {volume} {712}},\ \bibinfo {pages} {319} (\bibinfo {year}
  {2012})}\BibitemShut {NoStop}%
\bibitem [{\citenamefont {Escher}\ \emph {et~al.}(2012)\citenamefont {Escher},
  \citenamefont {Burke}, \citenamefont {Dietrich}, \citenamefont {Scielzo},
  \citenamefont {Thompson},\ and\ \citenamefont {Younes}}]{escher2012compound}%
  \BibitemOpen
  \bibfield  {author} {\bibinfo {author} {\bibfnamefont {J.~E.}\ \bibnamefont
  {Escher}}, \bibinfo {author} {\bibfnamefont {J.~T.}\ \bibnamefont {Burke}},
  \bibinfo {author} {\bibfnamefont {F.~S.}\ \bibnamefont {Dietrich}}, \bibinfo
  {author} {\bibfnamefont {N.~D.}\ \bibnamefont {Scielzo}}, \bibinfo {author}
  {\bibfnamefont {I.~J.}\ \bibnamefont {Thompson}}, \ and\ \bibinfo {author}
  {\bibfnamefont {W.}~\bibnamefont {Younes}},\ }\href@noop {} {\bibfield
  {journal} {\bibinfo  {journal} {Reviews of modern physics}\ }\textbf
  {\bibinfo {volume} {84}},\ \bibinfo {pages} {353} (\bibinfo {year}
  {2012})}\BibitemShut {NoStop}%
\bibitem [{\citenamefont {Hauser}\ and\ \citenamefont
  {Feshbach}(1952)}]{hauser1952inelastic}%
  \BibitemOpen
  \bibfield  {author} {\bibinfo {author} {\bibfnamefont {W.}~\bibnamefont
  {Hauser}}\ and\ \bibinfo {author} {\bibfnamefont {H.}~\bibnamefont
  {Feshbach}},\ }\href@noop {} {\bibfield  {journal} {\bibinfo  {journal}
  {Physical review}\ }\textbf {\bibinfo {volume} {87}},\ \bibinfo {pages} {366}
  (\bibinfo {year} {1952})}\BibitemShut {NoStop}%
\end{thebibliography}%
% Produces the bibliography via BibTeX.

\end{document}